\documentclass{article}

\usepackage[margin=1.2in]{geometry}
\usepackage{amsmath,amssymb,amsfonts,graphicx,amsthm,nicefrac,mathtools,bm}
\usepackage{hyperref}
\usepackage[numbers]{natbib}
\usepackage[english]{babel}
\usepackage{times}
\usepackage[T1]{fontenc}
\usepackage{bbm}
\usepackage{color,xcolor}
\usepackage{xspace}
\usepackage{rotating,multirow}

\newtheorem{theorem}{Theorem}

\newtheorem{lemma}{Lemma}
\newtheorem{proposition}{Proposition}

\theoremstyle{definition}
\newtheorem{definition}{Definition}
\theoremstyle{remark}

\makeatletter
\newtheorem*{rep@theorem}{\rep@title}
\newcommand{\newreptheorem}[2]{%
\newenvironment{rep#1}[1]{%
 \def\rep@title{#2 \ref{##1}}%
 \begin{rep@theorem}}%
 {\end{rep@theorem}}}
\makeatother
\newreptheorem{theorem}{Theorem}
\newreptheorem{lemma}{Lemma}
\newreptheorem{corollary}{Corollary}
\newreptheorem{proposition}{Proposition}

\newcommand{\figref}[1]{Figure~\ref{fig:#1}}

\newcommand{\secref}[1]{Section~\ref{sec:#1}}
\newcommand{\appref}[1]{Appendix~\ref{app:#1}}

\newcommand{\lemref}[1]{Lemma~\ref{lem:#1}}

\newcommand{\propref}[1]{Proposition~\ref{prop:#1}}

\newcommand{\thmref}[1]{Theorem~\ref{thm:#1}}

\newcommand{\thmsref}[1]{Theorems~\ref{thm:#1}}
\newcommand{\thmssref}[1]{\ref{thm:#1}}

\newcommand{\eqnref}[1]{\eqref{eqn:#1}}

\DeclareMathOperator{\diag}{diag}

\DeclareMathOperator{\var}{Var}
\DeclareMathOperator{\sign}{sign}

\newcommand{\iid}[0]{i.i.d.\xspace}

\newcommand{\One}[1]{{\mathbbm{1}}\left\{{#1}\right\}}
\newcommand{\inner}[2]{\langle{#1},{#2}\rangle} 
\newcommand{\norm}[1]{\lVert{#1}\rVert}

\newcommand{\PP}[1]{\mathbb{P}\left\{{#1}\right\}} 
\newcommand{\Pp}[2]{\mathbb{P}_{#1}\left\{{#2}\right\}} 
\newcommand{\EE}[1]{\mathbb{E}\left[{#1}\right]} 

\newcommand{\EEst}[2]{\mathbb{E}\left[{#1}\ \middle| \ {#2}\right]} 
\newcommand{\PPst}[2]{\mathbb{P}\left\{{#1}\ \middle| \ {#2}\right\}} 
\newcommand{\VV}[1]{\var\left({#1}\right)} 

\def\R{\mathbb{R}}

\newcommand{\ident}{\mathbf{I}}

\def\independenT#1#2{\mathrel{\rlap{$#1#2$}\mkern2mu{#1#2}}}
\newcommand\independent{\protect\mathpalette{\protect\independenT}{\perp}}
\newcommand{\iidsim}{\stackrel{\mathrm{iid}}{\sim}}

\newcommand{\ignore}[1]{}

\newcommand{\eps}{\epsilon}

\let\emptyset\varnothing

\newcommand{\PX}{P_X}
\newcommand{\PYX}{P_{Y\mid X}}
\newcommand{\Pj}{P_j}
\newcommand{\Phj}{Q_j}
\newcommand{\PhX}{Q_X}
\newcommand{\PXnotj}{P_{X_{-j}}}
\newcommand{\Pt}{P_{\bigxko\mid X}}
\newcommand{\X}{\mathbf{X}}
\newcommand{\xx}{\mathbf{x}}
\newcommand{\xxko}{\widetilde{\xx}}
\newcommand{\Xko}{\widetilde{\X}}
\newcommand{\xko}{\widetilde{x}}
\newcommand{\Zko}{\widetilde{Z}}
\newcommand{\bigxko}{\widetilde{X}}
\DeclareMathOperator{\swap}{swap}
\newcommand{\eqd}{\stackrel{\textnormal{d}}{=}}
\newcommand{\Scal}{\mathcal{S}}
\newcommand{\Shcal}{\widehat{\Scal}}

\newcommand{\fdr}{\textnormal{FDR}}
\newcommand{\y}{\mathbf{Y}}
\newcommand{\nulls}{\mathcal{H}_0}
\newcommand{\kl}{\widehat{\textnormal{KL}}}
\newcommand{\Thetah}{\widetilde{\Theta}}
\newcommand{\normal}{\mathcal{N}}
\newcommand{\PPround}[1]{\mathbb{P}\left({#1}\right)} 
\newcommand{\Ecal}{\mathcal{E}}
\usepackage{tikz}
\usetikzlibrary{calc}

\newcommand{\suppmat}[1]{{#1}}

\author{Rina Foygel Barber\thanks{Department of Statistics, University of Chicago}\quad\quad  
Emmanuel J.~Cand{\`e}s\thanks{Departments of Mathematics and of Statistics, Stanford University} \quad\quad 
Richard J.~Samworth\thanks{Statistical Laboratory, University of Cambridge}}
\title{Robust inference with knockoffs}

\date{February, 2019}

\begin{document}
\maketitle

\begin{abstract}

  We consider the variable selection problem, which seeks to identify
  important variables influencing a response $Y$ out of many candidate
  features $X_1, \ldots, X_p$. We wish to do so while offering
  finite-sample guarantees about the fraction of false
  positives---selected variables $X_j$ that in fact have no effect on
  $Y$ after the other features are known.  When the number of features
  $p$ is large (perhaps even larger than the sample size $n$), and we
  have no prior knowledge regarding the type of dependence between $Y$
  and $X$, the model-X knockoffs framework nonetheless allows us to
  select a model with a guaranteed bound on the false discovery rate,
  as long as the distribution of the feature vector
  $X=(X_1,\dots,X_p)$ is exactly known. This model selection procedure
  operates by constructing ``knockoff copies'' of each of the $p$
  features, which are then used as a control group to ensure that the
  model selection algorithm is not choosing too many irrelevant
  features.  In this work, we study the practical setting where the
  distribution of $X$ can only be estimated, rather than known
  exactly, and the knockoff copies of the $X_j$'s are therefore
  constructed somewhat incorrectly.  Our results, which are free of
  any modeling assumption whatsoever, show that the resulting model
  selection procedure incurs an inflation of the false discovery rate
  that is proportional to our errors in estimating the distribution of
  each feature $X_j$ conditional on the remaining features
  $\{X_k:k\neq j\}$.  The model-X knockoffs framework is therefore
robust to errors in the underlying assumptions on the distribution of
$X$, making it an effective method for many practical applications,
such as genome-wide association studies, where the underlying
distribution on the features $X_1,\dots,X_p$ is estimated accurately
but not known exactly.
\end{abstract}

\section{Introduction}\label{sec:intro}

Our methods of data acquisition are such that we often obtain
information on an exhaustive collection of possible explanatory
variables. We know a priori that a large proportion of these are
irrelevant for our purposes, but in an effort to cover all bases, we
gather data on all what we can measure and rely on subsequent analysis
to identify the relevant variables. For instance, to achieve a better
understanding of biological processes behind a disease, 
we may evaluate variation across the entire DNA sequence and collect
single nucleotide polymorphism (SNP) information, or quantify the
expression level of all genes, or consider a large panel of exposures,
and so on.  We then expect the statistician or the scientist to sort
through all these and select those important variables that truly
influence a response of interest. For example, we would like the
statistician to tell us which of the many genetic variations affect
the risk of a specific disease, or which of the many gene expression
profiles help determine the severity of a tumor. 

This paper is about this variable selection problem.  We consider
situations where we have observations on a response $Y$ and a large
collection of variables $X_1, \ldots, X_p$. With the goal of
identifying the important variables, we want to recover the smallest
set $\Scal \subseteq \{1, \ldots, p\}$ such that, conditionally on
$\{X_j\}_{j \in \Scal}$, the response $Y$ is independent of all the
remaining variables $\{X_j\}_{j\not\in \Scal}$. In the literature on
graphical models, the set $\Scal$ would be called the {\em
  Markov blanket} of $Y$.  Effectively, this means that the explanatory
variables $X_1,\dots,X_p$ provide information about the outcome  $Y$ only through the
subset $\{X_j\}_{j \in \Scal}$. To ensure reproducibility, we
are interested in methods that result in the estimation of a set
$\Shcal$ with false discovery rate  (FDR) control
\citep{BH95}, in the sense that
\[\fdr = 
\EE{\frac{\#\{ j : j \in \Shcal 
\setminus \Scal\}}{\#\{j: j \in \Shcal\}}}\leq q,
\]
i.e.~a bound on the expected proportion of our discoveries $\Shcal$
which are {\em not} in the smallest explanatory set
$\Scal$.\footnote{As is standard in the FDR literature, in this
  expected value we treat $0/0$ as $0$, to incur no penalty in the
  event that no variables are selected, i.e.~when $\Shcal=\emptyset$.}
(Here $q$ is some predetermined target error rate, e.g.~$q=0.1$.)

In truth, there are not many variable selection methods that would
control the FDR with finite-sample guarantees, especially when the
number $p$ of variables far exceeds the sample size $n$. That said,
one solution is provided by the recent model-X knockoffs approach of
\citet{candes2018panning}, which is a new read on the earlier knockoff
filter of \citet{barber2015}; see also \cite{barber2016}. One singular
aspect of the method of model-X knockoffs is that it makes assumptions
that are substantially different from those commonly encountered in
the statistical literature. Most of the model selection literature
relies on a specification of the model that links together the
response and the covariates, making assumptions on $\PYX$, the
distribution of $Y$ conditional on $X$---for instance, assuming that
the form of this distribution follows a generalized linear model or
some other parametrized model.  In contrast, model-X knockoffs makes
no assumption whatsoever on the relation between the response $Y$ and
the variables $X = (X_1, \ldots, X_p)$; in other words, the
distribution $\PYX$ of $Y$ conditional on $X$ is ``model free''.  The
price of this generality is that we need to be able to specify the
distribution of the feature variables $X = (X_1, \ldots, X_p)$, which
we denote by $\PX$.  This distribution is then used to construct knockoff feature 
variables $\bigxko = (\bigxko_1,\dots,\bigxko_p)$, where each $\bigxko_j$ mimics the real feature $X_j$
and acts a ``negative control'' in any variable selection algorithm---if our variable selection algorithm selects
any of the knockoff features, this alerts us to a high false positive rate in the algorithm.
Knowledge of the distribution of $X$ is needed in order to construct the $\bigxko_j$'s appropriately---for instance, 
if $X_1$ is a real signal while $X_2$ is null, then we need $\bigxko_2$ to mimic $X_2$'s dependence with $X_1$
in order to act as an appropriate negative control.

As argued in \cite{candes2018panning} and \cite{JansonPhD}, this
``shift'' of the burden of knowledge is interesting because we must
recall that the object of inference is on how $Y$ relates to $X$, that is, on $\PYX$. It
is, therefore, a strong premise to posit the form of this
relationship $\PYX$ a priori---and indeed, there are many applications in which we
objectively do not have any understanding of how $Y$ depends on
$X$. Further, the shift is also appropriate whenever we know much
more about the distribution of $X$ than on the conditional
distribution of $Y \, | \, X$.  For instance, it is easy to imagine
applications in which we have many unlabeled samples---samples of
$X$---whereas it may be much harder to acquire labeled data or samples
with a given value of the response $Y$. A typical example is offered
by genetic studies, where we now have available hundreds of thousands
or even millions of genotypes across many different populations. At the
same time, it may be difficult to recruit patients with a given
phenotype (the response variable $Y$), and therefore, we have substantially more data with which to estimate $\PX$ than
$\PYX$.

The ease with which we can gather information about $X$ does not imply
that we know the distribution $\PX$ exactly, but we often do
 have substantial information about this
distribution. Returning to our genetic example, it has been shown that
the joint distribution of SNPs may be accurately modeled  by hidden Markov
models (see \cite{SetD01,ZetS02,QetL02,Li2003} for some early
formulations), and there certainly is an abundance of genotype data to
estimate the various model parameters; compare for instance the
success of a variety of methods for genotype imputation \citep{MH10,
  HetA12} based on such models. 
  More generally, if a large amount of unlabeled data is available, the ``deep knockoffs'' methodology of \cite{romano2018deep}
 proposes using a deep generative model to generate knockoffs, subject
 to constraints that ensure that the knockoffs have approximately replicated the dependencies among the $X_j$'s.
 Empirically, they find that this method is extremely effective
at producing knockoff distributions that successfully control the FDR.

The purpose of this paper, then, is
precisely to investigate common situations of this kind, namely, what
happens when we run model-X knockoffs and only assume {\em approximate}
knowledge of the distribution of $X$ rather than exact knowledge, or equivalently, a construction
of the knockoff features $\bigxko$ that only {\em approximately} replicates the distribution of $X$.
Our contribution is a considerable extension of the original work on
model-X knockoffs \citep{candes2018panning}, which assumed a perfect
knowledge of the distribution of $X$ to achieve FDR control. 
If we only have access to an approximation of the distribution of $X$, then it is certainly
possible for model-X knockoffs to fail to control FDR---for instance, see \cite[Sec.~6.5,6.6]{romano2018deep} for 
examples where estimating the distribution of $X$ using only its first two moments is not sufficient for FDR control if the true distribution is heavy-tailed.

 Here, we develop a new theory, which quantifies very precisely the inflation in
FDR when running the knockoff filter with estimates of the
distribution of $X$ in place of the true distribution $\PX$. We
develop non-asymptotic bounds which show that the possible FDR
inflation is well-behaved whenever the estimated distribution is
reasonably close to the truth. These bounds are general and apply to
all possible statistics that the researcher may want to use to tease out
the signal from the noise. We also develop converse results for some
settings, showing that our bounds are fairly sharp in that it is
impossible to obtain tighter FDR control bounds in full generality.
Thus, our theory offers finite-sample guarantees that hold for any algorithm
that the analyst decides to employ, assuming no knowledge of the form of the relationship between $Y$ and $X$ and only
an estimate of the distribution of $X$ itself.  On the other hand, since our bounds are worst-case, they
  may be pessimistic in the sense that the realized FDR in any
  practical situation may be much lower than that achieved in the
  worst possible case.

Underlying our novel model-X knockoffs theory is a completely new
mathematical analysis and understanding of the knockoffs inferential
machine. The technical innovation here is essentially twofold. First,
with only partial knowledge of the distribution of $X$, we can no
longer achieve a perfect exchangeability between the test statistics
for the null variables and for their knockoffs. Hence, we need tools
that can deal with only a form of approximate exchangeability. Second,
our methods to prove FDR control no longer rely on martingale
arguments, and rather, involve leave-one-out type of arguments. These
new arguments are likely to have applications far outside the scope of
the present paper.

\section{Robust inference with knockoffs}

To begin with, imagine we have data consisting of $n$ \iid~draws from
a joint distribution on $(X,Y)$, where
$X = (X_1, \ldots, X_p) \in\R^p$ is the feature vector while $Y\in\R$
is the response variable.  We will gather the $n$ observed data points
into a matrix $\X\in\R^{n\times p}$ and vector $\y\in\R^n$---that is,
the pairs $(\X_{i,*},\y_i)$ are \iid~copies of the pair $(X,Y)$.  The
joint distribution of $(X,Y)$ is unknown---specifically, we do not
assume any information about the conditional distribution of $Y$ given
$X$ as discussed above.  We work under the assumption that $\PX$, the marginal distribution of $X$, is
known only approximately.

Since the Markov blanket of $Y$ may be ill-defined 
(e.g.~if two features are identical 
then the choice of the minimal set $\Scal$  may not be unique), we follow
\cite{candes2018panning} and define $X_j$ to be a null variable if
$X_j\independent Y\mid X_{-j}$, that is if $X_j$ and the response $Y$ are 
independent conditionally on all the other variables. 
(We use the
  terms ``features'' and ``variables'' interchangeably.) Under
 very
mild identifiability conditions, the set of non-nulls is nothing other than the Markov
blanket of $Y$. Writing $\nulls$ to denote the set of indices corresponding
to null variables, we can then  reformulate the error we would like to control
as
$\EE{|\Shcal\cap \nulls| / |\Shcal|} \leq q.$

\subsection{Exact model-X knockoffs}
\label{sec:exact}

Consider first an ideal setting where the distribution $\PX$ is known.
The model-X knockoffs method \citep{candes2018panning} is defined by constructing knockoff
features satisfying the following conditions: $\bigxko$ is drawn
conditional on the feature vector $X$ without looking at the response
$Y$ (i.e.~$\bigxko\independent Y\mid X$), such that the joint distribution of $(X,\bigxko)$ satisfies a
pairwise exchangeability condition, namely
\begin{equation}\label{eqn:swap_distribution}(X,\, \bigxko)_{\swap({\mathcal{A}})}
  \eqd (X,\, \bigxko)\end{equation}
for any subset ${\mathcal{A}}\subseteq\{1,\dots,p\}$, where $\eqd$ denotes
equality in distribution. (In fact, to achieve FDR control, this condition only needs to hold for
subsets ${\mathcal{A}}\subseteq\nulls$ containing only null variables.) Above, the family
$(X, \, \bigxko)_{\swap({\mathcal{A}})}$ is obtained from $(X, \, \bigxko)$ by swapping the
entries $X_j$ and $\bigxko_j$ for each $j\in {\mathcal{A}}$; for example, with $p = 3$
and ${\mathcal{A}} = \{2,3\}$,
\[
(X_1, X_2, X_3, \bigxko_1, \bigxko_2, \bigxko_3)
_{\swap{(\{2,3\})}} \, = \, (X_1, \bigxko_2, \bigxko_3,
\bigxko_1, X_2, X_3).
\]

As a consequence of the pairwise exchangeability property~\eqnref{swap_distribution}, 
we see that the null knockoff variables $\{\bigxko_j\}_{j \in \nulls}$
are distributed in exactly the same way as the original nulls
$\{X_j\}_{j \in \nulls}$ but some dependence is preserved: for
instance, for any pair $j\neq k$ where $k$ is a null, we have that
$(X_j, \bigxko_k) \eqd (X_j, X_k)$.

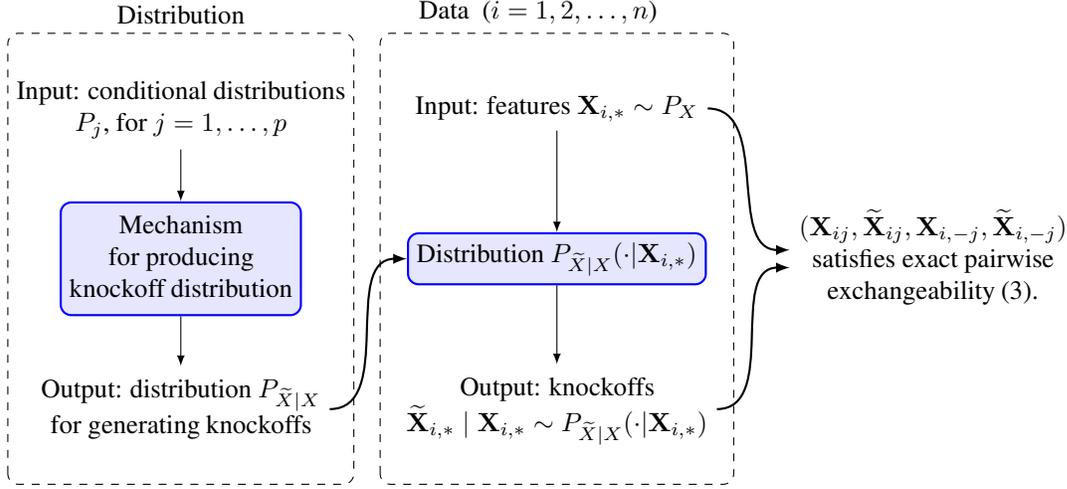
\begin{figure}[t] 
\begin{tikzpicture}

\node (inputconditionals) at (0,10)
{\begin{tabular}{@{}c@{}}Input: conditional distributions \\$\Pj$, for $j=1,\dots,p$\end{tabular}};
\node (mechanism) at (0,8)  [rectangle, rounded corners, text centered, draw=blue, fill=blue!10, thick]
  {\begin{tabular}{@{}c@{}}Mechanism\\ for producing\\ knockoff distribution\end{tabular}};
\node (outputPt) at (0,6)
{\begin{tabular}{@{}c@{}}Output: distribution $\Pt$\\ for generating knockoffs\end{tabular}};
\draw[-latex] (inputconditionals.south) to (mechanism.north);
\draw[-latex] (mechanism.south) to (outputPt.north);
\draw[rounded corners,dashed] (-2.3,5) rectangle (2.3,11) node[above] at (0,11) {Distribution};

\node (data) at (5,10)
{Input: features $\X_{i,*}\sim \PX$};
\node (usePt) at (5,8) [rectangle, rounded corners, text centered, draw=blue, fill=blue!10, thick]
 {Distribution $\Pt(\cdot|\X_{i,*})$};
\node (knockoffs) at (5,6)
{\begin{tabular}{@{}c@{}}Output:  knockoffs\\ $\Xko_{i,*}\mid \X_{i,*} \sim \Pt(\cdot | \X_{i,*})$\end{tabular}};
\draw[-latex] (data.south) to (usePt.north);
\draw[-latex] (usePt.south) to (knockoffs.north);
\draw[rounded corners,dashed] (2.65,5) rectangle (7.35,11) node[above] at (4.75,11) {Data \ ($i=1,2,\dots,n$)};

\node (pairwise) at (10,8) {\begin{tabular}{@{}c@{}}$(\X_{ij},\Xko_{ij},\X_{i,-j},\Xko_{i,-j})$\\ satisfies exact pairwise\\ exchangeability~\eqnref{swap_j_only}.\end{tabular}};

\draw[-latex,thick] (outputPt.east) to [in=180,out=0] (usePt.west);
\draw[-latex,thick] (data.east) to [in=180,out=0] ($(pairwise.west)!0.15!(pairwise.north west)$);
\draw[-latex,thick] (knockoffs.east) to [in=180,out=0] ($(pairwise.west)!0.15!(pairwise.south west)$);

\end{tikzpicture}
\caption{Schematic representation of the exact model-X knockoffs construction.  }
\label{fig:knockoff-mechanism-exact}
\end{figure}

Given knowledge of the true distribution $\PX$ of the features $X$,
our first step to implement the method of model-X knockoffs is to
construct a distribution for drawing $\bigxko$ conditional on $X$ such
that the pairwise exchangeability property~\eqnref{swap_distribution}
holds for all subsets of features ${\mathcal{A}}$.  We can think of
this mechanism as constructing some probability distribution
$\Pt(\cdot|x)$, which is a conditional distribution of $\bigxko$ given
$X=x$, chosen so that the resulting joint distribution of
$(X,\bigxko)$, which is equal to \[\PX(x) \, \Pt(\xko | x),\]
is symmetric in the pairs $(x_j,\xko_j)$, and thus will satisfy the
exchangeability property~\eqnref{swap_distribution}.  Now, when
working with data $(\X,\y)$, we will treat each data point $(\X_{i,*},\y_i)$
independently. Specifically, after observing
the data $(\X,\y)\in\R^{n\times p}\times \R^n$, the rows $\Xko_{i,*}$
of the knockoff matrix are drawn from $\Pt(\cdot|\X_{i,*})$,
independently for each $i$ and also independently of $\y$.
\figref{knockoff-mechanism-exact} shows a schematic representation of
the exact model-X knockoffs construction.

It is important to point out that mechanisms for producing the
pairwise exchangeability property~\eqnref{swap_distribution} do exist
and can be very concrete. As a specific example, suppose we wish to sample knockoff copies of Gaussian features,
which follow a known Gaussian distribution $\PX =  \normal_p(\mathbf{0}_p,\Sigma)$.
Then \citet{candes2018panning}  show that the knockoffs $\Xko_{i,*}$ can be drawn from the conditional distribution 
\begin{equation}\label{eqn:gaussian_knockoffs}\Pt(\cdot|\X_{i,*}) = \normal_p\big((\ident_p - D\Sigma^{-1})\X_{i,*},2D - D\Sigma^{-1}D\big)\end{equation}
for any fixed diagonal matrix $D$ satisfying $0\preceq D \preceq 2\Sigma$. (This mechanism provides valid knockoffs
because it ensures that the joint distribution of $(\X_{i,*},\Xko_{i,*})$ is given by
\[\normal_{2p}\left(\mathbf{0}_{2p}, \left(\begin{array}{@{}c@{\;\;}c@{}}\Sigma & \Sigma-D \\ \Sigma - D & \Sigma\end{array}\right)\right),\]
which satisfies pairwise exchangeability~\eqnref{swap_distribution}.)
There are also fast algorithms for the case where $X$ follows
either a Markov or a hidden Markov model \citep{Sesia2018gene}. 
More broadly, \citet{candes2018panning} develop a general
abstract mechanism termed the Sequential Conditional Independent Pairs
(SCIP) algorithm, which always produces exchangeable knockoff
copies and can be applied to any distribution $\PX$. Looking ahead, all of these algorithms can be used in the case
where $\PX$ is known only approximately, where the exchangeability property~\eqnref{swap_distribution}
will be required to hold only with reference to the {\em estimated} distribution of $X$,
discussed in \secref{approximate} below.

For assessing a model selection algorithm, the knockoff feature vectors
$\Xko_j$ can be used as a ``negative control''---a control group for testing the algorithm's
ability to screen out false positives, since $\Xko_j$ is known to have no real effect on $\y$. Although details are given in
\secref{filter_Wjs}, it is helpful to build some intuition already at
this stage. Imagine for simplicity that we wish to assess the
importance of a variable by measuring the strength of the marginal
correlation with the response, i.e.~we compute $Z_j = \big|\X_j^\top
\y\big|$. Then we can compare $Z_j$ with $\Zko_j = \big|\Xko_j^\top \y\big|$,
the marginal correlation for the corresponding knockoff variable. The
crucial point is that the pairwise exchangeability
property~\eqnref{swap_distribution} implies that if $j$ is null
(recall that this means that $X_j$ and $Y$ are conditionally
independent given $X_{-j}$), then
\[
(Z_j, \, \Zko_j) \eqd (\Zko_j, \, Z_j). 
\]
This holds without any assumptions on the form of the relationship $\PYX$
between $Y$ and $X$ \citep{candes2018panning}.  In particular, this 
means that the test statistic $W_j = Z_j - \Zko_j$ is equally likely 
to be positive or negative. Thus to reject the null, we would need to 
observe a large positive value of $W_j$. As we will see 
in \secref{filter_Wjs}, this way of reasoning extends to any choice of 
statistic $Z_j$; whatever statistic we choose, knockoff variables obeying~\eqnref{swap_distribution} 
offer corresponding values of the statistic which can be used as 
``negative controls'' for calibration purposes.

Throughout this paper, we will pay close attention to the distribution we
obtain when swapping only one variable and its knockoff (and do not
swap any of the other variables). In this context, we can reformulate the
broad exchangeability condition~\eqnref{swap_distribution} in terms of
single variable swaps.
\begin{proposition}[{\citet[Prop.~3.5]{candes2018panning}}]\label{prop:swap_j_only}
  The pairwise exchangeability property~\eqnref{swap_distribution}
  holds for a subset ${\mathcal{A}}\subseteq\{1,\dots,p\}$ if and only if
\begin{equation}\label{eqn:swap_j_only}
\big(X_j,\bigxko_j,X_{-j},\bigxko_{-j}\big) \eqd \big(\bigxko_j,X_j,X_{-j},\bigxko_{-j}\big)
\end{equation}
holds for all $j\in {\mathcal{A}}$. 
\end{proposition}
\noindent In other words, we can restrict our attention to the question of whether a single given feature $X_j$ and its knockoff $\bigxko_j$ are
exchangeable with each other (in the joint distribution that also includes $X_{-j}$ and $\bigxko_{-j}$).

\subsection{Approximate model-X knockoffs and pairwise
  exchangeability}
\label{sec:approximate}

Now we will work towards constructing a version of this method when
the true distribution $\PX$ of the feature vector $X$ is not known
exactly. Here, we need to relax the pairwise exchangeability
assumption, since choosing a useful mechanism $\Pt$ that satisfies
this condition would generally require a very detailed knowledge of
$\PX$, which is typically not available. This section builds towards
a definition of pairwise exchangeability with respect to an approximate estimate of $\PX$,
in two steps. 

\subsubsection{Exchangeability with respect to an input distribution $\PhX$}

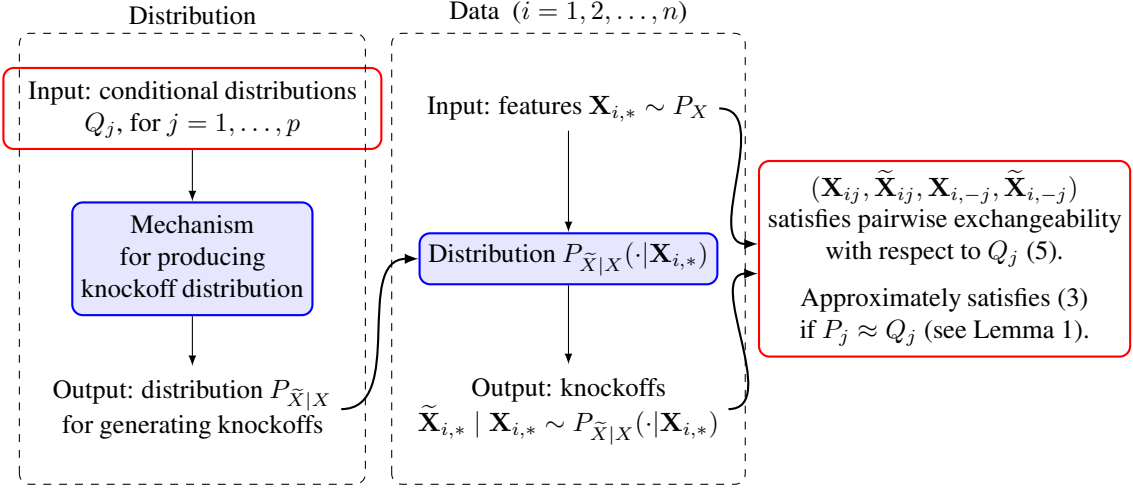
\begin{figure}[t] 
\begin{tikzpicture}

\node (inputconditionals) at (0,10) 
[rectangle, rounded corners, thick, text centered,draw=red]
{\begin{tabular}{c}Input: conditional distributions \\$\Phj$, for $j=1,\dots,p$\end{tabular}};
\node (mechanism) at (0,8)  [rectangle, rounded corners, text centered, draw=blue, fill=blue!10, thick]
  {\begin{tabular}{@{}c@{}}Mechanism\\ for producing\\ knockoff distribution\end{tabular}};
\node (outputPt) at (0,6) 
{\begin{tabular}{@{}c@{}}Output: distribution $\Pt$\\ for generating knockoffs\end{tabular}};
\draw[-latex] (inputconditionals.south) to (mechanism.north);
\draw[-latex] (mechanism.south) to (outputPt.north);
\draw[rounded corners,dashed] (-2.3,5) rectangle (2.3,11) node[above] at (0,11) {Distribution};

\node (data) at (5,10)
{Input: features $\X_{i,*}\sim \PX$};
\node (usePt) at (5,8) [rectangle, rounded corners, text centered, draw=blue, fill=blue!10, thick]
 {Distribution $\Pt(\cdot|\X_{i,*})$};
\node (knockoffs) at (5,6)  
{\begin{tabular}{@{}c@{}}Output:  knockoffs\\ $\Xko_{i,*}\mid \X_{i,*} \sim \Pt(\cdot | \X_{i,*})$\end{tabular}};
\draw[-latex] (data.south) to (usePt.north);
\draw[-latex] (usePt.south) to (knockoffs.north);
\draw[rounded corners,dashed] (2.65,5) rectangle (7.35,11) node[above] at (5,11) {Data \ ($i=1,2,\dots,n$)};

\node (pairwise) at (10,8) 
[rectangle, rounded corners, thick, text centered,draw=red]
{\begin{tabular}{@{}c@{}} $(\X_{ij},\Xko_{ij},\X_{i,-j},\Xko_{i,-j})$\\ satisfies pairwise exchangeability\\ with respect to  $\Phj$~\eqnref{swap_j_only_approx}.\medskip\\ 
Approximately satisfies~\eqnref{swap_j_only}\\ if $\Pj\approx \Phj$ (see \lemref{approx_pairwise}).\end{tabular}};

\draw[-latex,thick] (outputPt.east) to [in=180,out=0] (usePt.west);
\draw[-latex,thick] (data.east) to [in=180,out=0] ($(pairwise.west)!0.15!(pairwise.north west)$);
\draw[-latex,thick] (knockoffs.east) to [in=180,out=0] ($(pairwise.west)!0.15!(pairwise.south west)$);

\end{tikzpicture}
\caption{Schematic representation of the approximate model-X knockoffs
construction. The two differences relative to \figref{knockoff-mechanism-exact}
are circled in red.}
\label{fig:knockoff-mechanism-approx}
\end{figure}

We are provided with data $\X$ and conditional distributions
$\Phj(\cdot | x_{-j})$ for each $j$. We assume that the $\Phj$'s are
fixed, i.e.~do not depend on the data set $(\X,\y)$.  As a warm-up,
assume first that these conditionals are mutually compatible in the
sense that there is a joint distribution $\PhX$ over $\R^p$ that
matches these $p$ estimated conditionals---we will relax this
assumption very soon.
Then as shown in~\figref{knockoff-mechanism-approx}, we repeat the
construction from~\figref{knockoff-mechanism-exact}, only with the
$\Phj$'s as inputs. In words, the algorithm constructs knockoffs,
which are samples from $\Pt$, a conditional distribution whose
construction is based on the conditionals $\Phj$ or, equivalently, the
joint distribution $\PhX$.  In place of requiring that pairwise
exchangeability of the features $X_j$ and their knockoffs $\bigxko_j$
holds relative to the true distribution $P_X$ as
in~\eqnref{swap_distribution} and~\eqnref{swap_j_only}, we instead
require that the knockoff construction mechanism satisfy pairwise
exchangeability conditions relative to the joint distribution $\PhX$
that it receives as input:
\begin{equation}\label{eqn:swap_all_approx}
  \begin{aligned}
    \text{If $(X,\bigxko)$ is drawn as $X\sim \PhX$}
    \text{ and $\bigxko\mid X \sim \Pt(\cdot|X)$, then}\quad\quad\\
 \quad\quad (X,\, \bigxko)_{\swap({\mathcal{A}})}
  \eqd (X,\, \bigxko), \quad \text{for any subset
    ${\mathcal{A}}\subseteq\{1,\dots,p\}$}. 
\end{aligned}
\end{equation}
When only estimated compatible conditionals are available, original
and knockoff features are required to be exchangeable with respect to
the distribution $\PhX$, which is provided as input (but not with respect
 to the true distribution of $X$, which is unknown).
To rephrase, if the distribution of $X$ were in fact equal to $\PhX$,
then we would have exchangeability.

\subsubsection{Exchangeability with respect to potentially
  incompatible conditionals $\Phj$}

We wish to provide an extension of \eqnref{swap_all_approx} to
  cover the case where the conditionals may not be compatible; that
  is, when a joint distribution with the $\Phj$'s as conditionals may
  not exist.  To understand why this is of interest, imagine we have
  unlabeled data that we can use to estimate the distribution of
  $X$. Then we may construct $\Phj$ by regressing the $j$th feature
  $X_j$ onto the $p-1$ remaining features $X_{-j}$. For instance, we
  may use a regression technique promoting sparsity or some other
  assumed structure. In such a case, it is easy to imagine that such a
  strategy may produce incompatible conditionals. It is, therefore,
  important to develop a framework adapted to this setting. To address
  this, we shall work throughout the paper with the following
  definition:
\begin{definition}\label{def:exch_Qj}
$\Pt$ is {\em pairwise exchangeable with respect to $\Phj$}
if it satisfies the following property:
\begin{equation}\label{eqn:swap_j_only_approx}
  \begin{aligned}
    &\text{For any distribution $D^{(j)}$ on $\R^p$ with $j$th conditional $\Phj$, if $(X,\bigxko)$ is drawn as $X\sim D^{(j)}$ }\\
    &\text{\quad\quad and $\bigxko\mid X \sim \Pt(\cdot|X)$, then }\big(X_j,\bigxko_j,X_{-j},\bigxko_{-j}\big) \eqd
    \big(\bigxko_j,X_j,X_{-j},\bigxko_{-j}\big).
\end{aligned}
\end{equation}
Above, $D^{(j)}$ is the product of an arbitrary marginal distribution for
$X_{-j}$ and of the conditional $\Phj$. 
\end{definition}
\noindent In words, with estimated conditionals $\Phj$, we choose
$\Pt$ to satisfy pairwise exchangeability with respect to these
$\Phj$'s, for every $j$.  (As before, we remark that this only needs
to hold for $j\in\nulls$ to ensure FDR control, but since in practice
we do not know which features are null, we
require~\eqnref{swap_j_only_approx} to hold for every $j$.)

To see why this is an extension of \eqnref{swap_all_approx}, note that
if the $\Phj$'s are mutually compatible, i.e.~there is some distribution $\PhX$ with conditionals $\Phj$ for each $j$, then any algorithm operating
such that~\eqnref{swap_j_only_approx} holds for each $j$, obeys
\eqnref{swap_all_approx} as well---this is because, for each $j$, we can apply~\eqnref{swap_j_only_approx}
with the distribution $D^{(j)} = \PhX$. 

Now let's consider the question of how we might generate knockoff copies
obeying~\eqnref{swap_j_only_approx}. In the setting where our estimated conditionals $\Phj$
are all compatible with some
joint distribution $\PhX$ on $X$, constructing knockoff copies in this approximate scenario is no different
from the exact model-X knockoffs framework---if we have some mechanism which, when we input the joint distribution 
$\PX$ of $X$, will produce
exchangeable knockoffs obeying~\eqnref{swap_distribution},
then we can instead provide our estimated joint distribution $\PhX$ as input to produce knockoff copies 
that satisfy~\eqnref{swap_all_approx} and, by extension, satisfy~\eqnref{swap_j_only_approx}.
Hence, if the $\Phj$'s are mutually compatible, then
all the mechanisms producing valid knockoffs under exact knowledge of
$\PX$---we mentioned a few in the previous section---can be readily
used for our purposes. Later in \secref{examples}, we will also
give an example of a mechanism producing valid knockoffs
satisfying~\eqnref{swap_j_only_approx} under incompatible $\Phj$'s. 


\subsubsection{Probability of a swap}

We next develop a key lemma that will allow us to characterize the quality of our constructed knockoffs.
In an exact model-X knockoffs framework, the key idea is that the knockoffs $\bigxko_j$ act as controls for null variables
$X_j$, because even after all observing all of the data---all the covariates, and the response $Y$---we are
unable to tell which of the two, i.e.~$X_j$ and $\bigxko_j$, is the real variable versus the knockoff. More precisely,
each of the two is {\em equally likely} to be the real variable or the knockoff.
Our next step in the approximate setting, therefore, is to determine whether this is approximately true when the 
estimated conditionals $\Phj$ are not too far from the true conditionals $\Pj$.

From this point on, we will assume without comment that for each $j$, either $X_j$ and $\bigxko_j$
are both discrete variables or are both continuous variables,
and abusing notation, in these two settings we will use $\Pj(\cdot|x_{-j})$ and $\Phj(\cdot|x_{-j})$ to denote
the conditional probability mass function or conditional density, respectively, for the true and estimated
conditional distribution of $X_j$ given $X_{-j}=x_{-j}$. 
Furthermore, we assume that  $\Pj(\cdot|x_{-j})$ and $\Phj(\cdot|x_{-j})$ are supported on the same
(discrete or continuous) set for any $x_{-j}$. Our theory can be generalized to the
setting of mixed distributions and/or varying supports, but for clarity of the results we do
not present these generalizations here.

The construction of the knockoff features as in \figref{knockoff-mechanism-approx} yields the following approximate pairwise
exchangeability result (proved in \appref{proofs}).
\begin{lemma}\label{lem:approx_pairwise}
Fix any feature index $j$ such that pairwise exchangeability~\eqnref{swap_j_only_approx} with respect to $\Phj$ is satisfied.
If $X_j,\bigxko_j$ are discrete, then for any\footnote{Formally, this result holds only for $a,b$ lying in the support of 
$\Pj (\cdot | X_{-j})$, which is assumed to be equal to the support of $\Phj (\cdot | X_{-j})$, as otherwise the ratio is $0/0$; we ignore
this possibility here and throughout the paper since these results will be applied only in settings where $a,b$ do lie in this support.}
 $a,b$,
  \[\frac{\PPst{X_j = a,\bigxko_j =
      b}{X_{-j},\bigxko_{-j}}}{\PPst{X_j
      = b,\bigxko_j =
      a}{X_{-j},\bigxko_{-j}}}
  = \frac{\Pj (a | X_{-j})\Phj(b | X_{-j})}{\Phj(a | X_{-j}) \Pj (b |
    X_{-j})}.\] 
Furthermore, if index $j$ corresponds to a null feature (i.e.~$X_j\independent Y\mid X_{-j}$)
and we additionally assume that $\bigxko\mid X$ is drawn from
  $\Pt$ independently of $Y$, then the same result holds when we also condition on $Y$: 
 \begin{equation}\label{eqn:ratio_approx}
 \frac{\PPst{X_j = a,\bigxko_j =
      b}{X_{-j},\bigxko_{-j},Y}}{\PPst{X_j
      = b,\bigxko_j =
      a}{X_{-j},\bigxko_{-j},Y}}
  = \frac{\Pj (a | X_{-j})\Phj(b | X_{-j})}{\Phj(a | X_{-j}) \Pj (b |
    X_{-j})}.\end{equation} 
The conclusion in the continuous case is identical except with ratios of probabilities replaced with ratios of
densities.
\end{lemma}

To better understand the roles of the various distributions at play,
consider the two following scenarios for the joint distribution of the feature vector $X$ and its knockoff copy $\bigxko$:
\begin{figure}[h] 
\centering
\begin{tikzpicture}
\begin{scope}[xshift=0cm,yshift=0cm]
  \node [rotate=90] at (0,3) {True distribution}; \node at (2.5,3)
  {$\left\{\text{\begin{parbox}{1.5in}{
\begin{align*}
X_{-j}&\sim \text{(any distribution)}\\
X_j\mid X_{-j}&\sim \Pj(\cdot|X_{-j})\\
\bigxko\mid X &\sim \Pt(\cdot|X)\end{align*}
}\end{parbox}}\right.$};

\node [rotate=90] at (6,3) {Assumed distrib.};
\node at (8.5,3) {$\left\{\text{\begin{parbox}{1.5in}{
\begin{align*}
X_{-j}&\sim \text{(any distribution)}\\
X_j\mid X_{-j}&\sim \Phj(\cdot|X_{-j})\\
\bigxko\mid X &\sim \Pt(\cdot|X)\end{align*}
}\end{parbox}}\right.$};
\end{scope}
\end{tikzpicture}
\end{figure}

\noindent The knockoff generating mechanism $\Pt$ is designed with the estimated conditional $\Phj$ in mind, 
and therefore by construction, $X_j$ and $\bigxko_j$ are exchangeable under the ``Assumed distribution'' scenario on the right, 
defined with the incorrect estimate $\Phj$ of the $j$th conditional. The real distribution of $(X,\bigxko)$ instead follows the scenario labeled as the
``True distribution'', on the left. When $\Pj\neq \Phj$, this means
that $X_j$ and $\bigxko_j$ are only {\em approximately} exchangeable under the true distribution of the data.
\lemref{approx_pairwise} quantifies the extent to which the pair $(X_j,\bigxko_j)$ deviate from exchangeability, 
giving a useful formula for computing the ratio between the
likelihoods of the two configurations $(X_j,\bigxko_j)=(a,b)$ and
$(X_j,\bigxko_j)=(b,a)$ (after conditioning on the remaining data).

It is important to observe that if we are
working in the exact model-X framework, where  the true distribution and assumed distribution are the same (i.e.~$\Pj=\Phj$), 
then in this
case the lemma yields
\begin{equation}\label{eqn:ratio_exact}
\frac{\PPst{X_j = a,\bigxko_j = b}{X_{-j},\bigxko_{-j},Y}}{\PPst{X_j = b,\bigxko_j = a}{X_{-j},\bigxko_{-j},Y}} = 1\end{equation}
for each null $j$. That is, the two configurations are equally
likely. This result for the exact model-X setting is proved
in \citet[Lemma 3.2]{candes2018panning} and is critical for
establishing FDR control properties.
When we use
estimates $\Phj$ rather than the true conditionals $\Pj$, however,
the property~\eqnref{ratio_exact} is no longer true, since \lemref{approx_pairwise} shows that the ratio is 
no longer equal to $1$ in general. We can no longer use the knockoff
statistics as exact negative controls; only as approximate
controls. This is where the major difficulty comes in: if a knockoff
statistic is only approximately distributed like its corresponding
null, what is the potential inflation of the type-I error that this
could cause? In other words, if $\Phj\approx \Pj$ so that the ratio in~\eqnref{ratio_approx} is slightly different from 1,
how much might this inflate the resulting FDR? 

Before proceeding with this question, we first give some additional background on the knockoff filter,
to see how the knockoff variables $\Xko_j$ will be used to test our hypotheses. We will then return in \secref{fdr}
to the question of how errors in constructing the knockoffs can affect the resulting FDR.

 \subsection{The knockoff filter}\label{sec:filter_Wjs}

 After constructing the variables $\Xko_j$, we apply the knockoff
 filter to select important variables. We here quickly rehearse the
 main ingredients of this filter and refer the reader to
 \cite{barber2015} and \cite{candes2018panning} for additional
 details; our exposition borrows from \cite{barber2016}.
 Suppose that for each variable $\X_j$ (resp.~each knockoff
 variable $\Xko_j$), we compute a score statistic $Z_j$
 (resp.~$\Zko_j$), such that
 \begin{equation*}
\label{eq:Z}
(Z_1, \ldots, Z_p, \Zko_1, \ldots, \Zko_{p}) = z\big([\X, \,
\Xko], \, \y\big),
\end{equation*}
with the idea that $Z_j$ (resp.~$\Zko_{j}$) measures the
importance of $X_j$ (resp.~$\bigxko_j$) in explaining $Y$. Assume
that the scores are ``knockoff agnostic'' in the sense that {switching} a
variable with its knockoff simply {switches} the components of $Z$ in
the same way. This means that 
\begin{equation}\label{eqn:swap_z}
  z\big([\X, \, \Xko]_{\swap({\mathcal{A}})}, \, \y\big)  =   z\big([\X, \, \Xko], \, \y\big) _{\swap({\mathcal{A}})}\end{equation}
i.e.~swapping $X_1$ and $\bigxko_1$ before calculating $Z$ has the effect of swapping
$Z_1$ and $\Zko_1$, and similarly swapping $X_2$ and $\bigxko_2$ swaps $Z_2$
and $\Zko_2$, and so on. Here, we emphasize that $Z_j$ may be an
arbitrarily complicated statistic. For instance, it can be defined as
the absolute value of a lasso coefficient, or some random forest
feature importance statistic; or, we may fit both a lasso model and a
random forest, and choose whichever one
has the lowest cross-validated error.

These scores are then combined in a single importance statistic for
the variable $X_j$ as
\begin{equation*}
\label{eq:W}
W_j = f_j(Z_j, \Zko_j) \eqqcolon w_j\big([\X, \, \Xko], \, \y\big),
\end{equation*} 
where $f_j$ is any anti-symmetric function, meaning that
$f_j(v,u) = - f_j(u,v)$. As an example, we may have
$W_j = Z_j - \Zko_j$, where the $Z_j$'s and $\Zko_j$'s are the
magnitudes of regression coefficients estimated by the lasso at a
value of the regularization parameter given by cross-validation,
say. Again, any choice of anti-symmetric function $f_j$ and score
statistic $Z_j$, no matter how complicated, is allowed.  By definition,
the statistics $W_j$ obey the {\em flip-sign property}, which says
that swapping the $j$th variable with its knockoff has the effect of
changing the sign of $W_j$ (since, by~\eqnref{swap_z} above, if we
swap feature vectors $\X_j$ and $\Xko_j$ then $Z_j$ and $\Zko_{j}$ get
swapped):
\begin{equation}
\label{eq:antisymmetry}
w_j\big([\X, \, \Xko]_{\swap({\mathcal{A}})}, \, \y\big) = 
 \begin{cases} w_j\big([\X, \, \Xko],\y\big), 
& j\not\in {\mathcal{A}},\\  - w_j\big([\X, \, \Xko],\y\big), & j\in {\mathcal{A}}.
\end{cases}
\end{equation}
The $W_j$'s are the statistics that the knockoff filter will use. The idea
is that large positive values of $W_j$ provide evidence against the
hypothesis that the distribution of $Y$ is conditionally independent
of $X_j$, while in contrast, if $j\in\nulls$, then $W_j$ has a symmetric distribution
and, therefore, is equally likely to take on positive or negative
values.

In fact, it is equally valid for us to define
$W_j=w_j\big([\X, \, \Xko],\y\big)$ for any function $w_j$ satisfying
the flip-sign property~\eqnref{swap_z}, without passing through the
intermediate stage of defining $Z_j$'s and $\Zko_j$'s, and from this
point on we do not refer to the feature importance scores
$Z_j,\Zko_j$ in our theoretical results. However, for better
understanding of the intuition behind the method, we should continue
to think of $W_j$ as comparing the apparent importance of the feature
$\X_j$ versus its knockoff $\Xko_j$ for modeling the response $\y$.

Now that we have test statistics for each variable, we need a
selection rule. For the knockoff filter, we choose a threshold $T_0>0$
by setting\footnote{We want $T_0$ to be positive and the formal
    definition is that the minimum in \eqnref{knockoff_stoppingtime}
    is taken over all $t>0$ taking on values in the set
    $\{|W_1|, \ldots, |W_p|\}$.}
\begin{equation}\label{eqn:knockoff_stoppingtime}
  T_0 = \min\left\{t > 0 :
    \frac{\#\{j: W_j \leq - t\}}{\#\{j: W_j\geq  t\}}
    \leq  q \right\}\;,
\end{equation}
where $q$ is the target FDR level.  The output of the procedure is the
selected model $ \Shcal = \{j : W_j \geq T_0\}.$ In \cite{barber2015},
it is argued that the ratio appearing in the right-hand side of
\eqref{eqn:knockoff_stoppingtime} is an estimate of the false
discovery proportion (FDP) if we were to use the threshold $t$---this
is true because $\PP{W_j\geq t} = \PP{W_j\leq -t}$ for any null
feature $j\in\nulls$, and so we would roughly expect
\begin{equation}\label{eqn:fdphat_intuition}
\begin{aligned} 
  \text{(\# false positives at threshold $t$)} &= \#\{j\in\nulls: W_j
  \geq t\} \\
  & \approx \#\{j\in\nulls : W_j \leq - t\}\\
& \leq \#\{j: W_j
  \leq - t\},\end{aligned} \end{equation} that is, the numerator
in~\eqnref{knockoff_stoppingtime} is an (over)estimate of the number
of false positives selected at the threshold $t$.  Hence, the
selection rule can be interpreted as a step-up rule, stopping the first
time our estimate falls below our target level. A slightly more
conservative procedure, the knockoff+ filter, is given by incrementing
the number of negatives by one, replacing the threshold
in~\eqnref{knockoff_stoppingtime} with the choice
\begin{equation}
  \label{eqn:knockoffplus_stoppingtime}
  T_+ = \min\left\{t > 0 : \frac{1+\#\{j: W_j \leq - t\}}{\#\{j:
      W_j\geq t\}} \leq q \right\}\;,
\end{equation}
and setting $\Shcal = \{j: W_j\geq T_+\}$. Formalizing the intuition of our rough
calculation~\eqnref{fdphat_intuition}, the false discovery rate control properties of these two procedures
 are studied in \cite{barber2015} under an exact pairwise exchangeability setting.

\section{FDR control results}\label{sec:fdr}

\subsection{Measuring errors in the distribution}
If the knockoff features are generated using a mechanism designed to mimic the estimated conditionals
$\Phj$ rather than the true conditional distributions $\Pj $, when can we hope for error control?
Intuitively, if the conditional distributions $\Pj $ and $\Phj$ are similar,
 then we might hope that the knockoff feature $\Xko_j$ is a reasonably good control
group for the original feature $\X_j$.

In order to quantify this, we begin by measuring the discrepancy between the true conditional $\Pj$
and its estimate $\Phj$. Define the random variable
\begin{equation}\label{eqn:klj_define}
\kl_j\coloneqq  \sum_i \log\left(\frac{\Pj (\X_{ij}|\X_{i,-j})\cdot \Phj(\Xko_{ij}|\X_{i,-j})}{\Phj(\X_{ij}|\X_{i,-j})\cdot \Pj (\Xko_{ij}|\X_{i,-j})}\right),\end{equation}
where the notation $\kl_j$ suggests the KL divergence. In fact,
$\kl_j$ is the {\em observed} KL divergence between 
$(\X_j,\Xko_j,\X_{-j},\Xko_{-j})$ and $(\Xko_j,\X_j,\X_{-j},\Xko_{-j})$.
To prove this,
working in the discrete case for simplicity,
\lemref{approx_pairwise} tells us that
\begin{align*}
 \sum_i \log&\left(\frac{\Pj (\xx_{ij}|\xx_{i,-j})\cdot \Phj(\xxko_{ij}|\xx_{i,-j})}{\Phj(\xx_{ij}|\xx_{i,-j})\cdot \Pj (\xxko_{ij}|\xx_{i,-j})}\right) \\
&\hspace{2cm}=
 \log\left(\frac{\PP{(\X_j,\Xko_j,\X_{-j},\Xko_{-j})=(\xx_j,\xxko_j,\xx_{-j},\xxko_{-j})}}{\PP{(\Xko_j,\X_j,\X_{-j},\Xko_{-j})=(\xx_j,\xxko_j,\xx_{-j},\xxko_{-j})}}
 \right)
\end{align*}
for any $\xx_j,\xxko_j,\xx_{-j},\xxko_{-j}$. Therefore, we see that 
\[\EE{\kl_j}= \textnormal{d}_{\textnormal{KL}} \Big((\X_j,\Xko_j,\X_{-j},\Xko_{-j}) \, \big\| \, (\Xko_j,\X_j,\X_{-j},\Xko_{-j})\Big),\]
where $ \textnormal{d}_{\textnormal{KL}} $ is the usual KL divergence between distributions. 

In the exact model-X setting, where the knockoff construction
mechanism $\Pt$ satisfies the pairwise exchangeability
property~\eqnref{swap_distribution}, \propref{swap_j_only} immediately
implies that
$(\X_j,\Xko_j,\X_{-j},\Xko_{-j})\eqd(\Xko_j,\X_j,\X_{-j},\Xko_{-j})$
and, thus, $\EE{\kl_j}=0$---and in fact, since we are using the true conditionals $\Pj$, or in other words
$\Phj=\Pj$, we would have $\kl_j  = 0$ always. In the approximate model-X framework, we
can interpret $\kl_j$ as measuring the extent to which the pairwise
exchangeability property~\eqnref{swap_j_only} is violated for a
specific feature $j$. We will see in our results below that
controlling the $\kl_j$'s is sufficient to ensure control of the false
discovery rate for the approximate model-X knockoffs method.  More
precisely, we will be able to bound the false positives coming from
those null features which have small $\kl_j$. 

\subsection{FDR control guarantee}\label{sec:fdr_thm}
We now present our guarantee for robust error control with the model-X knockoffs filter. 
The proof of this theorem appears in \appref{proofs}.
\begin{theorem}\label{thm:fdr_robust}
  Under the definitions above, for any $\eps\geq0$, consider the null
  variables for which $\kl_j \le \eps$. If we use the knockoff+
  filter, then the fraction of the rejections that correspond to such
  nulls obeys
\begin{equation}
\label{eqn:FDReps}
\EE{\frac{\big|\{j: j\in\Shcal\cap \nulls\, \text{\em and } \,
    \kl_j\leq \eps\}\big|}{|\Shcal|\vee 1}} \leq q \cdot e^\eps.
\end{equation}
In particular, this implies that the false discovery rate is bounded as
\begin{equation}
\label{eqn:FDRbound}
\fdr\leq \min_{\eps\geq 0}\left\{q\cdot e^{\eps} +
  \PPround{\max_{j\in\nulls}\kl_j>\eps}\right\}.
\end{equation}
Similarly, for the knockoff filter, for any $\eps\geq 0$, a slightly modified
fraction of the rejections that correspond to nulls 
with $\kl_j \le \eps$ obeys
\[\EE{\frac{\big|\{j: j\in\Shcal\cap \nulls \, \text{\em and } \, \kl_j\leq \eps\}\big|}{|\Shcal| + q^{-1}} }\leq q \cdot e^\eps,\]
and therefore, we obtain a bound on a modified false discovery rate:
\[\EE{\frac{\big|\Shcal\cap\nulls\big|}{|\Shcal| + q^{-1}}}\leq \min_{\eps\geq 0}\left\{q\cdot e^{\eps} + \PPround{\max_{j\in\nulls}\kl_j>\eps}\right\}.\]
\end{theorem}
\noindent In \secref{examples}, we will see concrete
examples where $\max_{j=1,\dots,p}\kl_j$ is small with high probability, yielding a meaningful
result on FDR control.

It worth pausing to unpack our main result a little. Clearly, we
cannot hope to have error control over {\em all} nulls if we have done
a poor job in constructing some of their knockoff copies, because our
knockoff ``negative controls'' may be completely off. Having said this,
\eqnref{FDReps} tells us that that if we restrict our definition of
false positives to only those nulls for which we have a reasonable
``negative control'' via the knockoff construction, then the FDR is controlled.
Since we do not make any assumptions,
this type of result is all one can really hope for. In other
words, exact model-X knockoffs make the assumption that the knockoff
features provide exact controls for each null, thus ensuring
control of the false positives; our new result removes this
assumption, and provides a bound on the false positives when counting
only those nulls for which the corresponding knockoff feature provides
an approximate control.  

 In a similar fashion, imagine running a multiple
comparison procedure, e.g.~the Benjamini--Hochberg procedure, with
p-values that are not uniformly distributed under the null. Then in
such a situation, we cannot hope to achieve error control over {\em
  all} nulls if some of the null p-values follow grossly incorrect
distributions. However, we may still hope to achieve reasonable
control over those nulls for which the p-value is close to
uniform.

A noteworthy aspect of this result is that it makes no modeling
assumption whatsoever. Indeed, our FDR control guarantees hold in any
setting---no matter the relationship $\PYX$ between $Y$ and $X$, no
matter the distribution $\PX$ of the feature vector $X$, and no matter
the test statistics $W$ the data analyst has decided to employ (as
long as $W$ obeys the flip-sign condition). What the theorem says is
that when we use estimated conditionals $\Phj$, if the $\Phj$'s are close to the
true conditionals $\Pj$ in the sense that the quantities $\kl_j$ are
small, then the FDR is well under control. (In the ideal case where we
use the true conditionals, then $\kl_j = 0$ for all $j \in \nulls$,
and we automatically recover the FDR-control result from
\citet{candes2018panning}; that is, we get FDR control at the nominal
level $q$ since we can take $\epsilon = 0$.)

Finally, we close this section by emphasizing that the proof of
\thmref{fdr_robust} employs
arguments that are completely different from those one finds in the
existing literature on knockoffs.
We discuss the novelties in our techniques in \appref{proofs}.

\subsubsection{Is KL the right measure?}\label{sec:klj_or_Ej}

As mentioned above, our theorem applies to
any construction of the statistics $W$, including adversarial
constructions that might be chosen deliberately to try to detect the
differences between the $X_j$'s and the $\bigxko_j$'s. It is therefore expected that in any
practical scenario, the achieved FDR would be lower than that
suggested by our upper bounds. In practice,
$W$ would  be chosen to try to identify strong correlations
with $Y$, and we would not expect that this type of statistic is
worst-case in terms of finding discrepancies between the distributions
of $X_j$ and $\bigxko_j$. In fact, empirical studies
\citep{candes2018panning,Sesia2018gene} have already reported on the
robustness of model-X knockoffs vis-\`a-vis possibly large model
misspecifications when $W$ is chosen to identify a strong dependence
between $X$ and $Y$. 

Examining our result more closely, we can see that our theorem applies to any statistic $W$
 because the $\kl_j$'s measure our ability to distinguish
between each $\X_j$ and its knockoff copy $\Xko_j$, and therefore if the two are virtually indistinguishable
(i.e.~$\kl_j$ is small), then {\em any} importance statistic $W$ 
is almost equally likely to have $W_j>0$ or $W_j<0$ (as long as $W$ obeys the ``flip-sign'' property~\eqref{eq:antisymmetry}).
 In other words, if $\kl_j$ is low, then $\Xko_j$ provides
a high quality ``control group'' for the null $\X_j$, under any choice of $W$.
However, when we run the knockoff filter in practice, our statistics $W=(W_1,\dots,W_p)$
provide only a coarse summary of the data $\X,\Xko,\y$. Even if the $p$-dimensional vectors
$\X_j$ and $\Xko_j$ contain sufficient information for us to distinguish between the original null variable
and its knockoff (due to a poor approximation $\Phj$ of $\Pj$), it is likely that much of this information
is lost when we observe only $W$ instead of the full data.
Therefore, a small $\kl_j$ is sufficient, but by no means necessary, for FDR control---$\kl_j$ being small means that we are unable
to distinguish between a null and its knockoff when viewing the full data, while for FDR control we only need
to establish that the two are indistinguishable when viewing the statistics $W_1,\dots,W_p$.

To formalize this idea, suppose that we fix some choice of statistic $W$ (i.e.~a map from the data $(\X,\Xko,\y)$ to the statistic $W=(W_1,\dots,W_p)$).
Suppose that random variables $E_1,\dots,E_p$ satisfy the following property:
\begin{equation}\label{eqn:knockoff_quality}
\PPst{W_j>0,E_j\leq \eps}{|W_j|,W_{-j}}  
\leq e^\eps\cdot\PPst{W_j<0}{|W_j|,W_{-j}}\ \forall \ \eps\geq 0, \ j\in\nulls.
\end{equation}
(We would generally choose the $E_j$'s to be functions of $(\X,\Xko,\y)$,
and would then interpret the probability as being taken with respect to the joint distribution of the data $(\X,\Xko,\y)$.)
For each null $j$, if $E_j$ is low then this means that, if we are only given access to the statistic $W$ (rather than viewing
the full data), then we do not have much hope of distinguishing between the $j$th feature and its knockoff copy.
The following lemma verifies that the $\kl_j$'s satisfy this property universally, i.e.~for any choice of the feature importance statistic $W$.
\begin{lemma}\label{lem:check_kl}
For any choice of statistic $W$ that obeys the ``flip-sign'' property~\eqref{eq:antisymmetry},
the random variables $\kl_j$ defined in~\eqnref{klj_define} satisfy the property~\eqnref{knockoff_quality}.
\end{lemma}

We will now generalize our FDR control result, \thmref{fdr_robust},
to replace $\kl_j$ with any knockoff quality measure $E_j$ satisfying the property~\eqnref{knockoff_quality}.
The proof of this theorem, and the lemma above, appear in \appref{proofs}.
\begin{theorem}\label{thm:fdr_robust_generalize}
Under the definitions above, let $W$ be a statistic satisfying the ``flip-sign'' property~\eqref{eq:antisymmetry}.
Suppose that, for this choice of $W$, the random variables $E_1,\dots,E_p$ satisfy the property~\eqnref{knockoff_quality}, meaning
that they measure the quality of the knockoffs with respect to $W$.
Then the conclusions of \thmref{fdr_robust} hold
with $E_j$ in place of $\kl_j$ for each $j$.
\end{theorem}
In particular, if the statistic $W$ reveals much less information than the full data set $\X,\Xko,\y$, then 
it may be possible to construct $E_j$'s that are in general much lower than the $\kl_j$'s, thus yielding a tighter bound on FDR.
It remains to be seen whether, in specific settings for the distribution of the data, there are natural
examples of the statistic $W$ that are amenable to constructing tightly controlled $E_j$'s to yield
tighter bounds on the resulting FDR. We aim to explore this question in future work, but here we give one potential example.
Suppose that the statistic $W$ depends on the data $\X,\Xko,\y$ only through some coarse summary statistics, for example,
only through $\X^\top\y$ and $\Xko^\top\y$. In this setting, for any values $a,b\in\R$, define
\[E_j(a,b) = \log\left(\frac{\PPst{(\X_j^\top\y,\Xko_j^\top\y)=(a,b)}{\X_{-j},\Xko_{-j},\y}}{\PPst{(\X_j^\top\y,\Xko_j^\top\y)=(b,a)}{\X_{-j},\Xko_{-j},\y}}\right)\]
(where the numerator and denominator are interpreted as conditional probabilities or conditional densities, as appropriate).
We can then take
\[E_j = E_j(\X_j^\top\y,\Xko_j^\top\y)\]
and, by our assumption on $W$, we can verify that these $E_j$'s satisfy the desired property~\eqnref{knockoff_quality}.
Now, will $E_j$ yield a better bound on FDR? We
can see that $E_j$ measures the extent to which the one-dimensional random variables  $\X_j^\top\y$ and $\Xko_j^\top\y$ are distinguishable from each other, after observing
the remaining data, i.e.~$\X_{-j},\Xko_{-j},\y$. In contrast, $\kl_j$ measures the same question for the full $n$-dimensional random vectors $\X_j$ and $\Xko_j$,
and therefore will in general be much larger than $E_j$.

\subsection{A lower bound on FDR} 

 Next, we ask whether it is possible to prove a converse to \thmref{fdr_robust}, which guarantees
FDR control as long as the $\kl_j$'s are small.  We are interested in knowing
whether bounding the $\kl_j$'s is in fact necessary for FDR
control---or is it possible to achieve an FDR control guarantee even
when the $\kl_j$'s are large?  Of course, as discussed in \secref{klj_or_Ej},
for a predefined choice of the statistic $W$, the $\kl_j$'s may yield very conservative
results. Here, however, we are interested in determining whether the $\kl_j$'s are indeed the right
measure of FDR inflation when we are aiming for a result that is universal over {\em all} FDR control methods.

Theorem~\ref{thm:lowerbound} below proves that, if there
is a feature $j$ for which $\kl_j$ does not concentrate near zero,
then we can construct an honest model selection method that, when
assuming that the conditional distribution of $X_j\mid X_{-j}$ is
given by $\Phj$, {\em fails} to control FDR at the desired level if
the true conditional distribution is in fact $\Pj$. By ``honest'', we
mean that the model selection method would successfully control FDR at
level $q$ if $\Phj$ were the true conditional distribution.  Our
construction does not run a knockoff filter on the data; it is instead
a hypothesis testing based procedure, meaning that the $\kl_j$'s
govern whether it is possible to control FDR in a general
sense. Hence, our converse is information-theoretic in nature and not
specific to the knockoff filter. The proof of \thmref{lowerbound} is given in \suppmat{\appref{more_proofs}}.

\begin{theorem}\label{thm:lowerbound}
Fix any distribution $\PX$, any feature index $j$, and any estimated conditional distribution $\Phj$. 
Suppose that there exists a knockoff sampling mechanism $\Pt$ 
that is pairwise exchangeable with respect to $\Phj$~\eqnref{swap_j_only_approx},  such that
\[\PP{\kl_j \geq \eps} \geq c\]
for some $\eps,c>0$ when $(\X,\Xko)$ is drawn from $\PX\times\Pt$. Then there exists a conditional distribution $\PYX$, and a testing procedure
$\Shcal$ that maps data $(\X,\y)\in\R^{n\times p}\times \R^n$ to a selected set of features $\Shcal(\X,\y)\subseteq\{1,\dots,p\}$, such that:
\begin{itemize}
\item If the data points $(\X_{i,*},\y_i)$ are \iid~draws from the distribution $\PhX \times \PYX$, where $\PhX$ is any distribution
whose $j$th conditional is $\Phj$ (that is,~our estimated conditional distribution $\Phj$ for feature $X_j$ is correct), then
\[\fdr\big(\Shcal\big) = q.\]
\item On the other hand,
if the data points $(\X_{i,*},\y_i)$ are \iid~draws from the distribution $\PX\times \PYX$ (i.e.~our estimated conditional distribution $\Phj$ is not correct, as the true conditional distribution is $\Pj$), then
\[\fdr\big(\Shcal\big)  \geq q \big(1 + c (1 - e^{-\eps})\big).\]
\end{itemize}
\end{theorem}
\noindent For the last case (where $\PX$ is the true distribution), if
$c\approx 1$ (i.e.~$\kl_j\geq \eps$ with high probability) then
$\fdr\big(\Shcal\big)\gtrapprox q(2-e^{-\eps})$; when $\eps\approx 0$
is small, we have $2-e^{-\eps}\approx 1 + \eps \approx e^\eps$, which is the same inflation
factor on the FDR on the upper
bound in \thmref{fdr_robust}.
 In other words, \thmsref{fdr_robust} and \thmssref{lowerbound} provide
(nearly) matching upper and lower bounds. With these theorems, we do not aim to claim that the knockoffs methodology is 
universally robust, but rather, to determine and quantify the robustness properties of this already existing method. It is indeed
true that substantial mistakes in the model of $X$ can lead to a loss of FDR control, and 
the theorems above show that the $\kl_j$'s quantify exactly when, and to what extent, this issue has the potential to occur. 
Of course, as discussed above in \secref{klj_or_Ej}, if we restrict our attention to prespecified statistics $W$,
then the actual loss of FDR control maybe 
much less severe than that predicted by the bounds in \thmref{fdr_robust}.

\section{Examples}\label{sec:examples}

To make our FDR control results more concrete, we will give two examples of settings where
accurate estimates $\Phj$ of the conditionals $\Pj$ ensure that the $\kl_j$'s are bounded near zero.
Examining the definition~\eqnref{klj_define} of $\kl_j$, we see that $\kl_j$ is a sum of $n$ \iid~terms,
and we can therefore expect that large deviation bounds such as Hoeffding's inequality can be used 
to provide an upper bound uniformly across all $p$ features. (Of course, as noted in \secref{klj_or_Ej},
measuring knockoff quality via the $\kl_j$'s is a ``worst-case'' analysis that will bound FDR universally
over all statistics $W$, and may therefore give a very conservative result; for a specific predefined choice of $W$, 
it may be possible to compute a tighter bound.) 

All theoretical results in this section are proved in \suppmat{\appref{more_proofs}}.

\subsection{Bounded errors in the likelihood ratio}
First, suppose that our estimates $\Phj$ of the conditional distribution $\Pj$ satisfy
a likelihood ratio bound uniformly over any values for the variables:
\begin{equation}\label{eqn:delta_bound_wp1}
\log\left(\frac{\Pj (x_j\mid x_{-j})\cdot \Phj(x_j'\mid x_{-j})}{\Phj(x_j\mid x_{-j})\cdot \Pj (x_j'\mid x_{-j})} \right)\leq \delta\end{equation}
for all $j$, all $x_j,x'_j$, and  all $x_{-j}$.
In this setting, the following lemma,
 proved via Hoeffding's inequality, 
gives a  bound on the $\kl_j$'s:
\begin{lemma}\label{lem:boundederror}
If the condition~\eqnref{delta_bound_wp1} holds uniformly for all $j$ and all $x_j,x'_j,x_{-j}$,
then with probability at least $1-\frac{1}{p}$,
\[\max_{j=1,\dots,p}\kl_j \leq \frac{n\delta^2}{2} +  2\delta\sqrt{n\log(p)}   .\]
\end{lemma}
\noindent In other words, if $\Phj$ satisfies~\eqnref{delta_bound_wp1}
for some $\delta = o\left(\frac{1}{\sqrt{n\log(p)}}\right)$, then with high probability every $\kl_j$ will be small.
By \thmref{fdr_robust}, then, the FDR for model-X knockoffs in this setting is controlled near the target level~$q$.

\subsection{Gaussian knockoffs}
For a second example,
suppose that the  distribution of the feature vector $X$ is mean zero and has covariance $\Theta^{-1}$,
where $\Theta$ is some unknown precision matrix. (We assume zero mean for simplicity, but these results
can of course be generalized to an arbitrary mean.)
 Suppose that we have estimated  $\Theta$ with some approximation $\Thetah$, and let $\Theta_j$ and $\Thetah_j$ 
denote the $j$th columns of these matrices. Our results below will assume
that the error in estimating each column of $\Theta$ is small,
i.e.~$\Thetah_j-\Theta_j$ is small for all $j$. 

As described earlier in~\eqref{eqn:gaussian_knockoffs}, \citet[eqn.~(3.2)]{candes2018panning}'s Gaussian knockoff construction
consists of drawing the knockoffs according to the conditional distribution $\Pt(\cdot|X)$ given by
\begin{equation}\label{eqn:draw_knockoffs_Gaussian}
  \bigxko \mid X \sim \normal_p\big((\ident_p - D\Thetah)  X ,2D-D\Thetah D\big),\end{equation}
where $D=\diag\{d_j\}$ is a nonnegative diagonal matrix chosen to satisfy $2D - D\Thetah D\succeq 0$, or equivalently, $D\preceq 2\Thetah^{-1}$.
If the true precision matrix of $X$ were given by $\Thetah$ (assumed to be
positive definite), 
then we can calculate that the joint distribution of the pair $(X,\bigxko)$ has first and second moments given by
\[\EE{\left(\begin{array}{c} X \\ \bigxko\end{array}\right)} = \left(\begin{array}{c} 0 \\ 0 \end{array}\right), \quad 
\VV{\left(\begin{array}{c} X \\ \bigxko\end{array}\right)} = \left(\begin{array}{cc}
\Thetah^{-1} & \Thetah^{-1} - D \\ \Thetah^{-1} - D & \Thetah^{-1}\end{array}\right).\]
In other words, for every $j$, $X_j$ and $\bigxko_j$ are exchangeable if we only look at the first and second
moments of the joint distribution.

If the true distribution of $X$ is in fact Gaussian, again with mean
zero and covariance $\Thetah^{-1}$, then a stronger claim
follows---the joint distribution of $(X,\bigxko)$ is then multivariate
Gaussian and therefore
$(X,\, \bigxko)_{\swap({\mathcal{A}})} \eqd (X,\, \bigxko)$ for every
subset $\mathcal{A}\subseteq [p]$.  In other words, the knockoff
construction determined by $\Pt$ satisfies pairwise exchangeability,
as defined in~\eqnref{swap_all_approx}, with respect to the
distribution $\PhX = \normal_p(0,\Thetah^{-1})$.  To frame this
property in terms of conditionals, $\Phj$, rather than an estimated
joint distribution, $\PhX$, we can calculate the estimated conditional
distributions $\Phj(\cdot|X_{-j})$ as
\begin{equation}\label{eqn:gaussian_Qj}
X_j\mid X_{-j} \sim \normal\left(X_{-j}^\top \left( \frac{- \Thetah_{-j,j}}{  \Thetah_{jj}}\right), \frac{1}{ \Thetah_{jj}}\right),
\end{equation}
where $\Thetah_{-j,j}\in\R^{p-1}$ is the column $\Thetah_j$ with entry
$\Thetah_{jj}$ removed. 

As noted in \secref{approximate}, we may want to work with estimated
precision matrices, which are not positive semidefinite (PSD). The
rationale is that if $\Thetah$ is fitted by regressing each $X_j$ on
the remaining features $X_{-j}$ to produce the $j$th column,
$\Thetah_j$, then the result will not be PSD in general. If $\Thetah$
is not PSD, although there is no corresponding joint distribution, the
conditionals $\Phj$~\eqnref{gaussian_Qj} are still well-defined as
long as $\Thetah_{jj}>0$ for all $j$; they are just not
compatible. (Note that symmetry is a far easier constraint to enforce,
e.g.~by simply replacing our initial estimate $\Thetah$ with
$(\Thetah+\Thetah^\top)/2$, which preserves desirable features such as
sparsity that might be present in the initial $\Thetah$; in contrast,
projecting to the PSD cone while enforcing sparsity constraints may be
computationally challenging in high dimensions.)

Our first result verifies that this construction of $\Pt$  satisfies pairwise exchangeability with respect to the conditional distributions $\Phj$
given in~\eqnref{gaussian_Qj}:
\begin{lemma}\label{lem:gaussian_pairwise}
Let $\Thetah\in\R^{p\times p}$ be a symmetric matrix with  a positive diagonal, and let $\Pt$ be defined as in~\eqnref{draw_knockoffs_Gaussian}. 
Then, for each $j=1,\dots,p$, $\Pt$ is pairwise exchangeable with respect to the conditional distribution $\Phj$
given in~\eqnref{gaussian_Qj}---that is, the exchangeability condition~\eqnref{swap_j_only_approx} is satisfied.
\end{lemma}

In practice, we would construct Gaussian knockoffs in situations where
the distribution of $X$ might be well approximated by a multivariate
normal. The lemma below gives a high probability bound on the
$\kl_j$'s in the case where the features are indeed Gaussian but with
an unknown covariance matrix $\Theta^{-1}$.  Here, Gaussian
concentration results can be used to control the $\kl_j$'s, which then
yields FDR control. (We note that recent work by \citet{fan2017rank}
also studies the Gaussian model-X knockoffs procedure with an
estimated precision matrix $\Thetah$, under a different framework.)
\begin{lemma}\label{lem:Gaussian}
  Let $\Theta,\Thetah\in\R^{p\times p}$ be any matrices, where
  $\Theta$ is positive definite and $\Thetah$ is symmetric with a
  positive diagonal.  Suppose that
    $\X_{i,*}\iidsim \normal_p(0,\Theta^{-1})$, while $\Xko\mid \X$ is
    drawn according to the distribution $\Pt$ given
    in~\eqnref{draw_knockoffs_Gaussian}. Define
\begin{equation}\label{eqn:delta_Theta}\delta_{\Theta} = \max_{j=1,\dots,p} (\Theta_{jj})^{-1/2}\cdot \norm{\Theta^{-1/2}(\Thetah_j -\Theta_j)}_2.\end{equation}
Then with probability at least $1-\frac{1}{p}$,
\[\max_{j=1,\dots,p}\kl_j \leq  4\delta_\Theta\sqrt{n\log(p)}\cdot (1+o(1)),\]
where the $o(1)$ term refers to terms that are vanishing when we assume that $\frac{\log(p)}{n}=o(1)$
and that this upper bound is itself bounded by a constant.
\end{lemma}
\noindent (A formal bound making the $o(1)$ term explicit is provided in the proof.)
In particular, comparing to our FDR control result, \thmref{fdr_robust}, we see that
as long as the columnwise error in estimating the precision matrix $\Theta$ satisfies
$\delta_\Theta = o\left(\frac{1}{\sqrt{n\log(p)}}\right)$, the FDR
will be controlled near the target level $q$.

When might we be able to attain such a bound on the error in estimating $\Theta$?
As mentioned earlier, in many applied settings, we may have access to substantially more unlabeled data (i.e.~the
feature vector $X$ without an associated response $Y$) than labeled data (pairs $(X,Y)$). Suppose that, for the purpose of estimating
$\Theta$, we have access to $N\gg n$ draws of the feature vector $X\sim\PX$. When the distribution
of $X$ is multivariate Gaussian with a sparse inverse covariance matrix $\Theta$, 
 the graphical Lasso \citep{yuan2007model,friedman2008sparse} estimates $\Theta$ as
\[\widehat{\Theta}_{\lambda} = \arg\min_{A\succeq 0} \biggl\{ - \log\det(A) + \inner{A}{\widehat{S}_N} + \lambda\sum_{j\neq k}|A_{jk}|\biggr\},\]
where $\widehat{S}_N$ is the sample covariance matrix of the unlabeled training data while $\lambda>0$ is a penalty parameter inducing sparsity in the resulting solution.
\citet{ravikumar2011high} proved that, if $\Theta$ is sufficiently sparse, then
 under certain additional assumptions and with an appropriate choice of penalty parameter $\lambda$,
 the graphical Lasso solution $\widehat{\Theta}_{\lambda}$
satisfies an entrywise error bound $\norm{\widehat\Theta_{\lambda} - \Theta}_{\infty}\lesssim \sqrt{\frac{\log(p)}{N}}$,
and furthermore, is asymptotically guaranteed to avoid any false positives (i.e.~if $\Theta_{jk}=0$ then $(\widehat\Theta_\lambda)_{jk}=0$).
Therefore, if each column of $\Theta$ has at sparsity at most $s_\Theta$ (i.e.~at most $s_\Theta$ nonzeros) and $\Theta$ has bounded
condition number, this then proves that the bound~\eqnref{delta_Theta}
on the error in estimating $\Theta$ holds with $\delta_{\Theta}\asymp \sqrt{\frac{s_\Theta\log(p)}{N}}$.
We conclude that the results of \lemref{Gaussian} give a meaningful bound on FDR control as long as
\[ 4\delta_\Theta\cdot \sqrt{n\log(p)} \asymp \sqrt{\frac{s_\Theta\log(p)}{N}} \cdot \sqrt{n\log(p)} = o(1).\]
Equivalently, it is sufficient to have an unlabeled sample size $N$ satisfying
\[N \gg n\cdot s_\Theta\log^2(p).\]

\section{Discussion}

In this paper, we established that the method of model-X knockoffs is
robust to errors in the underlying assumptions on the distribution of
the feature vector $X$, making it an effective method for many
practical applications, such as genome-wide association studies, where
the underlying distribution on the features $X_1,\dots,X_p$ can be
estimated accurately. One notable aspect is that our theory is free of
any modeling assumptions, since our theoretical guarantees hold no
matter the data distribution or the statistics that the data analyst wishes
to use, even if they are designed to exploit some weakness in the
construction of knockoffs. Looking forward, it would be interesting to
develop a theory for fixed statistics, as outlined in \secref{klj_or_Ej}.
For instance, if the researcher
commits to using a pre-specified random forest feature importance
statistic, or some statistic based on the magnitudes of lasso
coefficients (perhaps calculated at a data-dependent value of the
regularization parameter), then what can be said about FDR control? In
other words, what can we say when the statistics $W$ only probe the
data in certain directions? We leave such interesting questions for
further research.

\subsection*{Acknowledgements}
R.~F.~B.~was partially supported by the National Science Foundation
via grant DMS 1654076, and by an Alfred P.~Sloan fellowship.
E.~C.~was partially supported by the Office of Naval Research under
grant N00014-16-1-2712, by the National Science Foundation via DMS
1712800, by the Math + X Award from the Simons Foundation and by a
generous gift from TwoSigma. E.~C.~would like to thank Chiara Sabatti
for useful conversations related to this project. R.~J.~S.~was
 partially supported by Engineering and Physical Sciences 
Research Council Fellowships EP/J017213/1 and EP/P031447/1, and by grant 
RG81761 from the Leverhulme Trust.

\bibliographystyle{plainnat}
\bibliography{robustknockoffsbib}

\appendix

\section{Proofs of main results}
\label{app:proofs}

Whereas all proofs of FDR control for the knockoff methods thus far
have relied on martingale arguments (see
\cite{barber2015,barber2016,candes2018panning}), here we will prove
our main theorem using a novel leave-one-out argument. Before we
begin, we would like to draw a loose analogy.  To prove FDR
controlling properties of the Benjamini--Hochberg procedure under
independence of the p-values, \citet{storey2004} developed a very
elegant martingale argument. Other proof techniques, however, operate
by removing or leaving out one hypothesis (or one p-value); see
\citet{BY01,ferreira2006} for examples. 
  At a very high level, our own methods are
partially inspired by the latter approach.

\subsection{Proofs of FDR control results, \thmsref{fdr_robust} and \thmssref{fdr_robust_generalize}}

\thmref{fdr_robust} follows directly from \thmref{fdr_robust_generalize} combined with \lemref{check_kl}, and thus
requires no separate proof. To prove \thmref{fdr_robust_generalize}, for any $\eps \geq 0$ and for any threshold $t > 0$, define
\[R_{\eps}(t) \coloneqq \frac{\sum_{j\in\nulls } \One{W_j\geq t,\kl_j\leq \eps}}{1 + \sum_{j\in\nulls} \One{W_j\leq -t}}.\]
Then, for the knockoff+ filter with threshold $T_+$, we can write
\begin{multline*}
 \frac{\big|\{j: j\in\Shcal\cap \nulls\text{ and }\kl_j\leq \eps\}\big|}{|\Shcal|\vee 1} 
   = \frac{\sum_{j\in\nulls } \One{W_j\geq T_+,\kl_j\leq \eps}}{1 \vee\sum_j \One{W_j\geq T_+}} \\
 = \frac{1+\sum_j \One{W_j\leq -T_+}}{1\vee \sum_j \One{W_j\geq T_+}}\cdot \frac{\sum_{j\in\nulls} \One{W_j\geq T_+,\kl_j\leq\eps}}{1 + \sum_j \One{W_j\leq -T_+}}\\
  \leq \frac{1+\sum_j \One{W_j\leq -T_+}}{1\vee \sum_j \One{W_j\geq T_+}}\cdot R_\eps(T_+) \leq q\cdot R_\eps(T_+),\end{multline*}
where the next-to-last step holds by definition of $R_{\eps}$, and the last step holds by the construction of the knockoff+ filter.
If we instead use the knockoff filter (rather than knockoff+), then we use the threshold $T_0$ and similarly obtain
\begin{align*}
\frac{\big|\{j: j\in\Shcal\cap \nulls\text{ and }\kl_j\leq \eps\}\big|}{q^{-1} + |\Shcal|} &\leq \frac{1 + \sum_j \One{W_j\leq -T_0}}{q^{-1}+\sum_j \One{W_j\geq T_0}}\cdot  R_\eps(T_0) \\
&\leq q\cdot R_\eps(T_0),
\end{align*}
where the two steps hold by definition of $R_{\eps}$ and the construction of the knockoff filter, respectively.
Either way, then, it is sufficient to prove that $\EE{R_{\eps}(T)}\leq e^\eps$, where $T$ is either $T_+$ or $T_0$.

Next, given a threshold rule $T=T(W)$ mapping statistics $W\in\R^p$ to a 
threshold $T> 0$ (i.e.~the knockoff or knockoff+ filter threshold, $T_0$ or $T_+$), for each index $j=1,\dots,p$ we define
\[T_j = T\Big((W_1,\dots,W_{j-1},|W_j|,W_{j+1},\dots,W_p)\Big) > 0,\]
i.e.~the threshold that we would obtain if $W_j$ were replaced with $|W_j|$. The following lemma (proved in \suppmat{\appref{more_proofs}})
 establishes a property of the $T_j$'s in the context of the knockoff filter:
\begin{lemma}\label{lem:Tj} Let $T=T(W)$ be the threshold for either the knockoff or the knockoff+.\footnote{More 
generally, this result holds for any function $T=T(W)$ that
satisfies a ``stopping time condition'' with respect to the signs of
the $W_j$'s, defined as follows: for any $t>0$, the event
$\One{T\leq t}$ depends on $W$ only through (1) the magnitudes $|W|$,
(2) $\sign(W_j)$ for each $j$ with $|W_j|<t$, and (3)
$\sum_{j:|W_j|\geq t} \sign(W_j)$.}
 For any $j,k$,
\begin{equation}\label{eqn:leave_one_out_threshold}
\text{If $W_j\leq -\min\{T_j,T_k\}$ and $W_k\leq -\min\{T_j,T_k\}$, then $T_j=T_k$.}\end{equation}
\end{lemma}

Now with $T$ being either the knockoff or knockoff+ thresholding rule,
we have
\begin{align*}
\EE{R_{\eps}(T)}
&=\EE{\frac{  \sum_{j\in\nulls}\One{W_j\geq T, E_j\leq \eps}}{1+\sum_{j\in\nulls} \One{W_j \leq -T}}} \\
&= \sum_{j\in\nulls}\EE{\frac{ \One{W_j\geq T_j, E_j\leq \eps}}{1+\sum_{k \in\nulls, k\neq j} \One{W_k \leq -T_j}}},
\end{align*}
where the last step holds since $T>0$ by definition, so if $W_j\geq T$ then $W_j\not\leq -T$, and, by definition of $T_j$, we also have $T=T_j$ in this case.
Continuing from this last step, we can rewrite the expectation as
\begin{align*}
\EE{R_{\eps}(T)}
&= \sum_{j\in\nulls}\EE{\frac{ \One{W_j>0, E_j\leq \eps}\cdot\One{|W_j|\geq T_j}}{1+\sum_{k\in\nulls, k\neq j}\One{W_k \leq -T_j}}}\\
&\stackrel{\text{(*)}}{=} \sum_{j\in\nulls}\EE{\frac{\PPst{W_j>0, E_j\leq \eps}{|W_j|,W_{-j}}\cdot\One{|W_j|\geq T_j}}{1+\sum_{k\in\nulls, k\neq j}\One{W_k \leq -T_j}}}\\
&\leq e^\eps\cdot  \sum_{j\in\nulls}\EE{\frac{\PPst{W_j<0}{|W_j|,W_{-j}}\cdot\One{|W_j|\geq T_j}}{1+\sum_{k\in\nulls, k\neq j}\One{W_k \leq -T_j}}} \\
&\stackrel{\text{(*)}}{=}  e^\eps\cdot  \sum_{j\in\nulls}\EE{\frac{ \One{W_j<0}\cdot\One{|W_j|\geq T_j}}{1+\sum_{k\in\nulls, k\neq j}\One{W_k \leq -T_j}}}\\
&= e^\eps\cdot \EE{ \sum_{j\in\nulls}\frac{ \One{W_j\leq  -T_j}}{1+\sum_{k\in\nulls, k\neq j}\One{W_k \leq -T_j}}},
\end{align*}
where the two steps marked with (*) hold because $T_j$ is a function of $|W_j|,W_{-j}$ by its definition, and so we
can treat it as known when we condition on $|W_j|,W_{-j}$.

Finally, the  summation inside the last expected value above can be simplified as follows:
if for all null $j$, $W_j > -T_j$, then the sum is equal to zero, while otherwise, we can write
\begin{multline*}\sum_{j\in \nulls} {\frac{ \One{W_j\leq -T_j}}{1+\sum_{k \in\nulls, k\neq j} \One{W_k \leq -T_j}}} 
= \sum_{j\in\nulls}\frac{\One{W_j\leq -T_j}}{1+\sum_{k\in\nulls,k\neq j}\One{W_k\leq -T_k}}\\
 = \sum_{j \in \nulls} {\frac{ \One{W_j\leq -T_j}}{\sum_{k\in\nulls} \One{W_k \leq -T_k}}} = 1,\end{multline*}
where the first step applies Lemma~\ref{lem:Tj}.
Combining everything, we have shown that
$\EE{R_{\eps}(T)}\leq e^\eps$, which proves
the theorem.

\subsection{Proof of \lemref{check_kl}}
We need to prove that
\[\PPst{W_j>0,\kl_j\leq \eps}{|W_j|,W_{-j}} \leq e^{\eps} \cdot \PPst{W_j<0}{|W_j|,W_{-j}}\]
for any null $j$ and any $\eps\geq 0$.
To proceed,  we will be conditioning on observing
$\X_{-j},\Xko_{-j},\y$, and on observing the {\em unordered} pair
$\{\X_j,\Xko_j\}$---that is, we observe both the original and knockoff
features but do not know which is which. It follows from the flip-sign
property that having observed all this, we know all the knockoff
statistics $W$ except for the sign of the $j$th component $W_j$. Put
differently, $W_{-j}$ and $|W_j|$ are both functions of the variables
we are conditioning on, but $\sign(W_j)$ is not. 
 Without loss of generality, label the
unordered pair of feature vectors $\{\X_j,\Xko_j\}$, as $\X^{(0)}_j$
and $\X^{(1)}_j$, such that:
\begin{equation}\label{eqn:X0_X1}\begin{cases}
\text{If $\X_j = \X^{(0)}_j$ and $\Xko_j = \X^{(1)}_j$, then $W_j\geq 0$;}\\
\text{If $\X_j = \X^{(1)}_j$ and $\Xko_j = \X^{(0)}_j$, then $W_j\leq 0$.}\end{cases}\end{equation}
We can therefore write
\begin{align*}
&\PPst{W_j>0,\kl_j\leq \eps}{|W_j|,W_{-j}}  \\
&\hspace{1cm}= \EEst{\PPst{W_j>0,\kl_j\leq \eps}{\X^{(0)}_j,\X^{(1)}_j,\X_{-j},\Xko_{-j},\y}}{|W_j|,W_{-j}}
\end{align*}
and similarly
\begin{align*}&\PPst{W_j<0}{|W_j|,W_{-j}} \\
&\hspace{2cm}= \EEst{\PPst{W_j<0}{\X^{(0)}_j,\X^{(1)}_j,\X_{-j},\Xko_{-j},\y}}{|W_j|,W_{-j}}.\end{align*}
Therefore, it will be sufficient to prove that
\begin{align}\label{eqn:kl_conditional}&\PPst{W_j>0,\kl_j\leq \eps}{\X^{(0)}_j,\X^{(1)}_j,\X_{-j},\Xko_{-j},\y} \nonumber \\
&\hspace{2cm}\leq e^\eps \cdot \PPst{W_j<0}{\X^{(0)}_j,\X^{(1)}_j,\X_{-j},\Xko_{-j},\y}.\end{align}
Now, if $\X^{(0)}_j,\X^{(1)}_j,\X_{-j},\Xko_{-j},\y$ are such that $|W_j|=0$, clearly this bound holds trivially, so from this point
on we ignore this trivial case and assume that $|W_j|>0$.
By our definition~\eqnref{X0_X1} of $\X^{(0)}_j$ and $\X^{(1)}_j$, we have
\begin{multline}\label{eqn:rho_j_step1}\quad\quad\frac{\PPst{W_j>0}{\X^{(0)}_j,\X^{(1)}_j,\X_{-j},\Xko_{-j},\y}}{\PPst{W_j<0}{\X^{(0)}_j,\X^{(1)}_j,\X_{-j},\Xko_{-j},\y}} \\= 
\frac{\PPst{(\X_j,\Xko_j)=(\X^{(0)}_j,\X^{(1)}_j)}{\X^{(0)}_j,\X^{(1)}_j,\X_{-j},\Xko_{-j},\y}}{\PPst{(\X_j,\Xko_j)=(\X^{(1)}_j,\X^{(0)}_j)}{\X^{(0)}_j,\X^{(1)}_j,\X_{-j},\Xko_{-j},\y}},\quad\quad\end{multline}
where this last ratio should be interpreted as a ratio of conditional probabilities or conditional densities, as appropriate.
Since the observations $i=1,\dots,n$ are independent, this can be rewritten as
\begin{multline}\label{eqn:rho_j_step2}\prod_{i=1}^n \frac{\PPst{(\X_{ij},\Xko_{ij})=(\X^{(0)}_{ij},\X^{(1)}_{ij})}{\X^{(0)}_{ij},\X^{(1)}_{ij},\X_{i,-j},\Xko_{i,-j},\y_i}}{\PPst{(\X_{ij},\Xko_{ij})=(\X^{(1)}_{ij},\X^{(0)}_{ij})}{\X^{(0)}_{ij},\X^{(1)}_{ij},\X_{i,-j},\Xko_{i,-j},\y_i}} \\
= \prod_{i=1}^n \frac{\Pj (\X^{(0)}_{ij}\mid \X_{i,-j})\cdot \Phj(\X^{(1)}_{ij}\mid\X_{i,-j})}{\Phj(\X^{(0)}_{ij}\mid \X_{i,-j})\cdot \Pj (\X^{(1)}_{ij}\mid\X_{i,-j})} \eqqcolon e^{\rho_j},\end{multline}
where the first equality holds by \lemref{approx_pairwise} (recalling that $j$ is assumed to be a null feature). Next, from the definition~\eqnref{klj_define} of $\kl_j$ and the definition~\eqnref{X0_X1} of $\X^{(0)}_j$ 
and $\X^{(1)}_j$, we can see that $\kl_j=\rho_j$ if $W_j>0$, or otherwise $\kl_j=-\rho_j$ if $W_j<0$.
Therefore,
\begin{multline*}\PPst{W_j>0,\kl_j\leq \eps}{\X^{(0)}_j,\X^{(1)}_j,\X_{-j},\Xko_{-j},\y}
 = \PPst{W_j>0,\rho_j\leq \eps}{\X^{(0)}_j,\X^{(1)}_j,\X_{-j},\Xko_{-j},\y} \\
= \One{\rho_j\leq \eps}\cdot \PPst{W_j>0}{\X^{(0)}_j,\X^{(1)}_j,\X_{-j},\Xko_{-j},\y}\\
= \One{\rho_j\leq \eps}\cdot e^{\rho_j} \cdot  \PPst{W_j<0}{\X^{(0)}_j,\X^{(1)}_j,\X_{-j},\Xko_{-j},\y},\end{multline*}
where the next-to-last step holds since $\rho_j$ is a function of $\X^{(0)}_j,\X^{(1)}_j,\X_{-j},\Xko_{-j},\y$,
while the last step uses our work in~\eqnref{rho_j_step1} and~\eqnref{rho_j_step2}.
Since $\One{\rho_j\leq \eps}\cdot e^{\rho_j}\leq e^\eps$ trivially, we have proved the desired bound~\eqnref{kl_conditional},
which concludes the proof of the lemma.

\subsection{Proof of \lemref{approx_pairwise}}
We prove the lemma in the case where all features are discrete; the case where some of the features may be continuous
is proved analogously. First, consider any null feature index $j$.  By definition of
the nulls, we know that $X_j\independent Y \mid X_{-j}$. Furthermore,
$\bigxko\independent Y \mid X$ by construction.  Therefore, the
distribution of $Y\mid (X,\bigxko)$ depends only on $X_{-j}$, and in
particular,
$Y\independent (X_j,\bigxko_j) \mid (X_{-j},\bigxko_{-j})$.
This proves that
\begin{equation}\label{eqn:ratio1} \frac{\PPst{X_j = a,\bigxko_j =
      b}{X_{-j},\bigxko_{-j},Y}}{\PPst{X_j
      = b,\bigxko_j =
      a}{X_{-j},\bigxko_{-j},Y}} = \frac{\PPst{X_j = a,\bigxko_j = b}{X_{-j},\bigxko_{-j}}}{\PPst{X_j = b,\bigxko_j = a}{X_{-j},\bigxko_{-j}}},\end{equation}
because the numerator and denominator are each unchanged whether we do or do not condition on $Y$.
Thus, for null features $j$, it is now sufficient to prove only the first claim of the lemma,
namely that the right-hand side above is equal to $\frac{\Pj (a | X_{-j})\Phj(b |X_{-j})}{\Phj(a | X_{-j}) \Pj (b |X_{-j})}$.

From this point on, let $j$ be any feature (null or non-null).
We will now prove the first claim in the lemma.
Recalling the assumption that $\Pt$ is pairwise exchangeable with respect to 
$\Phj$~\eqnref{swap_j_only_approx}, we
introduce a pair of random variables drawn as follows: first, draw
$X'_{-j}\sim \PXnotj$, where $\PXnotj$ is the distribution of $X_{-j}$;  then
draw  $X'_j\mid X'_{-j}\sim\Phj(\cdot|X'_{-j})$;
and finally, draw $\bigxko'\mid X'\sim \Pt(\cdot|X')$. Then by~\eqnref{swap_j_only_approx},
\begin{equation}\label{eqn:swap_j_only_approx_primes}
\big(X'_j,\bigxko'_j,X'_{-j},\bigxko'_{-j}\big) \eqd 
\big(\bigxko'_j,X'_j,X'_{-j},\bigxko'_{-j}\big) .\end{equation}
By construction, the joint distribution of $(X', \bigxko')$ is given by
\[
\PP{X' = x, \bigxko' = \xko} =  \PXnotj(x_{-j}) \Phj(x_j \, | x_{-j}) \Pt(\xko\,|\,x).
\]
Now, fixing any $x_{-j},\xko_{-j}\in\R^{p-1}$, write $x^a$ as the vector in $\R^p$ with entry $j$ given by $a$
and all other entries given by $x_{-j}$, and define $x^b,\xko^a,\xko^b$ analogously. 
Then~\eqnref{swap_j_only_approx_primes} is equivalent to
\begin{equation}\label{eqn:lem1_step}\PXnotj(x_{-j}) \Phj(a \, | x_{-j}) \Pt(\xko^b\,|\,x^a) = \PXnotj(x_{-j}) \Phj(b \, | x_{-j}) \Pt(\xko^a\,|\,x^b).\end{equation}

Now we turn to the true distribution of the data, generated as $X\sim \PX$ and $\bigxko\mid X\sim \Pt$. This means that
the joint distribution of $(X, \bigxko)$ is given by
\[
\PP{X = x, \bigxko = \xko} =  \PXnotj(x_{-j}) \Pj(x_j \, | x_{-j}) \Pt(\xko\,|\,x).
\]
We can therefore calculate
\begin{multline*}\frac{\PP{X_j=a,\bigxko_j=b,X_{-j}=x_{-j},\bigxko_{-j}=\xko_{-j}}}{\PP{X'_j=a,\bigxko'_j=b,X'_{-j}=x_{-j},\bigxko'_{-j}=\xko_{-j}}}\\
= \frac{\PXnotj(x_{-j}) \Pj(a \, | x_{-j}) \Pt(\xko^b\,|\,x^a)}{\PXnotj(x_{-j}) \Pj(b \, | x_{-j}) \Pt(\xko^a\,|\,x^b)}
=  \frac{\Pj(a\mid x_{-j})}{\Phj(a\mid x_{-j})}\cdot  \frac{\Phj(b\mid x_{-j})}{\Pj(b\mid x_{-j})},\end{multline*}
where the last step holds by~\eqnref{lem1_step}.
This proves the lemma.

\section{Additional proofs}\label{app:more_proofs}

\subsection{Proof of \thmref{lowerbound}}
First, we will show that our statement can be reduced to a binary hypothesis testing problem. 
We will work under the global null hypothesis where $Y\independent X$,
and our test will be constructed independently of $Y$. More formally, let $\PYX$ be any fixed
distribution, e.g.~$\normal(0,1)$. 
Since all features are null, this means that the false discovery proportion is $1$ whenever $\Shcal(\X,\y)\neq\emptyset$,
that is,
\[\fdr\big(\Shcal\big) = \PP{\Shcal(\X,\y)\neq \emptyset}.\]
Therefore, in order to prove the theorem, it is sufficient to construct a {\em binary}
test $\psi(\X)\in\{0,1\}$ such that 
\begin{equation}\label{eqn:psi_errors}
\Pp{\X_{i,*}\iidsim \PX}{\psi(\X) = 1}\geq  q\big(1+c(1-e^{-\eps})\big), \quad \Pp{\X_{i,*}\iidsim \PhX}{\psi(\X) = 1}= q ,\end{equation}
i.e.~a test $\psi$ that has better-than-random performance for testing whether the conditional distribution
of $X_j$ is given by $\Pj$ or $\Phj$. Once $\psi$ is constructed, then this is sufficient for the FDR result, e.g.~setting 
\[\Shcal(\X,\y) = \begin{cases} \{j\},&\psi(\X) = 1,\\ \emptyset,&\psi(\X)=0.\end{cases}\]
Note that, by the well-known equivalence between total variation distance and
hypothesis testing \citep{lehmann2008testing}, the existence of such a test $\psi$ is  essentially equivalent to 
proving a lower bound on
\[\textnormal{d}_{\textnormal{TV}}\left((\PX)^{\otimes n},\big(\PhX)^{\otimes n}\right)\]
uniformly over all distributions $\PhX$ whose $j$th conditional is $\Phj$.  In fact, our $\psi$
will be given by a randomized procedure (to be fully formal, we can
use the independent random vector $\y$ as a source of randomness, if
needed). First, we draw $\Xko\mid \X$, independently of $\y$ and drawn
from the rule $\Pt$ as specified in the theorem, and independently we
also draw $B\sim\textnormal{Bernoulli}(2q)$ and
$B'\sim\textnormal{Bernoulli}(q)$. Next, defining $\kl_j$ as in~\eqnref{klj_define},
 we let
\[\psi(\X,\Xko,B,B') = \One{B=1\text{ and } \kl_j >0 } + \One{B'=1\text{ and }\kl_j=0}.\]
Clearly, by definition of $B$ and $B'$, we have
 \begin{equation}\label{eqn:checking_psi}\PP{\psi(\X,\Xko,B,B') = 1} = 2q\cdot \PP{\kl_j>0} + q\cdot \PP{\kl_j=0},\end{equation}
where $\PP{\kl_j>0}$ and $\PP{\kl_j=0}$ are taken with respect to the joint distribution of $(\X,\Xko)$.

Next, we check that the test $\psi$ satisfies the
properties~\eqnref{psi_errors}, as required for the FDR bounds in this
theorem. We first prove the second bound in~\eqnref{psi_errors}.
Suppose $\X_{i,*}\iidsim \PhX$---that is, $\Phj$ is
indeed the correct conditional distribution for $X_j\mid X_{-j}$. The
knockoff generating mechanism $\Pt$ was defined to satisfy pairwise exchangeability with respect to
$\Phj$~\eqnref{swap_j_only_approx}, meaning that $\X_j$ and $\Xko_j$
are exchangeable conditional on the other variables in this scenario.
Examining the form of $\kl_j$, we see that swapping $\X_j$ and
$\Xko_j$ has the effect of changing the sign of $\kl_j$.  The
exchangeability of the pair $(\X_j,\Xko_j)$ implies that the
distribution of $\kl_j$ is symmetric around zero, and so under
$(\X_{i,*},\Xko_{i,*})\iidsim \PhX\times\Pt$, 
 \[\PP{\kl_j>0} +  0.5\cdot  \PP{\kl_j=0}=0.5.\]
 Checking~\eqnref{checking_psi}, this proves that
 $\Pp{\X_{i,*}\iidsim \PhX}{\psi(\X) = 1}= q$, which
 ensures FDR control for the case that the estimated conditional
 $\Phj$ is in fact correct.

Finally we turn to the first part of~\eqnref{psi_errors}, where now we assume that $(\X_{i,*},\Xko_{i,*})\iidsim \PX\times \Pt$.
From this point on, we will condition on the observed values of $\X_{-j}$ and $\Xko_{-j}$.
By assumption in the theorem, under this distribution we have
$\PP{\kl_j\geq \eps}\geq c$.
As in the proof of \lemref{check_kl},
we consider the unordered pair $\{\X_j,\Xko_j\}$---that is, we see the two vectors
$\X_j$ and $\Xko_j$ but do not know which is which. Note that, with this information,
we are able to compute $\big|\kl_j\big|$ but not $\sign(\kl_j)$.
Without loss of generality, we can label the
unordered pair of feature vectors $\{\X_j,\Xko_j\}$, as $\X^{(0)}_j$
and $\X^{(1)}_j$, such that
\begin{itemize}
\item if $\X_j = \X^{(0)}_j$ and $\Xko_j = \X^{(1)}_j$, then $\kl_j\geq 0$;
\item if $\X_j = \X^{(1)}_j$ and $\Xko_j = \X^{(0)}_j$, then $\kl_j\leq 0$.
\end{itemize}
Define $C=\sign(\kl_j)$, so that $\kl_j = C\cdot \big|\kl_j\big|$. 
By definition of the distribution of $(\X,\Xko)$,
 it follows from  \lemref{approx_pairwise}
that 
\[
\frac{\PPst{(\X_j,\Xko_j) =
    (\X^{(0)}_j,\X^{(1)}_j)}{\X^{(0)}_j,\X^{(1)}_j, \X_{-j}}}{\PPst{(\X_j,\Xko_j) =
    (\X^{(1)}_j,\X^{(0)}_j)}{\X^{(0)}_j,\X^{(1)}_j, \X_{-j}}} =
\prod_i  \frac{\Pj (\X^{(0)}_{ij}\mid\X_{i,-j})\Phj(\X^{(1)}_{ij}\mid
  \X_{i,-j})}{  \Phj(\X^{(0)}_{ij}\mid\X_{i,-j})\Pj (\X^{(1)}_{ij}\mid
  \X_{i,-j})}.  
\]
In other words, if $|\kl_j|\neq 0$, then
\[\frac{\PPst{C=+1}{\X^{(0)}_j,\X^{(1)}_j,\X_{-j},\Xko_{-j}}}{\PPst{C=-1}{\X^{(0)}_j,\X^{(1)}_j,\X_{-j},\Xko_{-j}}}
= \prod_i  \frac{\Pj (\X^{(0)}_{ij}\mid\X_{i,-j})\Phj(\X^{(1)}_{ij}\mid
  \X_{i,-j})}{  \Phj(\X^{(0)}_{ij}\mid\X_{i,-j})\Pj (\X^{(1)}_{ij}\mid
  \X_{i,-j})} \\
  = \exp\left\{\big|\kl_j\big|\right\},\]
where the last step holds by our choice of which vector to label as $\X^{(0)}$ and which to label as $\X^{(1)}$. 

Therefore, we can write
\begin{align}
\notag c\leq \PP{\kl_j\geq \eps} & = \PP{C = +1\text{ and }
  \big|\kl_j\big|\ge \eps}\\\notag & =
\EE{\PPst{C=+1}{\X^{(0)}_j,\X^{(1)}_j,\X_{-j},\Xko_{-j}}\cdot
  \One{\big|\kl_j\big|\geq \eps}} \\ \label{eqn:kl_prob_to_exp_eps}& =
\EE{\frac{e^{\big|\kl_j\big|}}{1+e^{\big|\kl_j\big|}}\cdot\One{\big|\kl_j\big|\geq
    \eps}}.\end{align}
We can similarly calculate
\[
\PP{\kl_j>0}
=\EE{\frac{e^{\big|\kl_j\big|}}{1+e^{\big|\kl_j\big|}}\cdot \One{\big|\kl_j\big|>0}}.
\]
Therefore,
\begin{multline*}
\frac{1}{2}\PP{\kl_j=0} + \PP{\kl_j>0} \\
 =\EE{\frac{e^0}{1+e^0}\cdot\One{\big|\kl_j\big|=0}} + \EE{\frac{e^{\big|\kl_j\big|}}{1+e^{\big|\kl_j\big|}}\cdot \One{\big|\kl_j\big|>0}}
=\EE{\frac{e^{\big|\kl_j\big|}}{1+e^{\big|\kl_j\big|}}}.
\end{multline*}
To continue, observe that for $t \ge 0$, $e^t/(1+e^t) \ge 1/2$. Hence,
\begin{align*}
\EE{\frac{e^{\big|\kl_j\big|}}{1+e^{\big|\kl_j\big|}}} &\geq \frac{1}{2} +\EE{ \left(\frac{e^{\big|\kl_j\big|}}{1+e^{\big|\kl_j\big|}} - \frac{1}{2}\right) \cdot \One{\big|\kl_j\big|\geq \eps}}\\
&\geq \frac{1}{2} +\min_{t\geq \eps} \frac{\frac{e^t}{1+e^t} - \frac{1}{2}}{\frac{e^t}{1+e^t}} \cdot \underbrace{\EE{\frac{e^{\big|\kl_j\big|}}{1+e^{\big|\kl_j\big|}}\cdot \One{\big|\kl_j\big|\geq \eps}}}_{\textnormal{$\geq c$ by~\eqnref{kl_prob_to_exp_eps}}}\\
&\geq \frac{1}{2}\left( 1 + c(1-e^{-\eps})\right),
\end{align*}
where for the last step we check that the minimum is attained at $t=\eps$. This proves that, when $\X_{i,*}\iidsim\PX$,
we have $\psi(\X,\Xko,B,B')=1$ with probability at least $q\big(1+c(1-e^{-\eps})\big)$, 
and so the first part of~\eqnref{psi_errors} is satisfied, as desired.

\subsection{Proof of \lemref{boundederror}}
We will in fact prove a more general result, which will be useful later on:
\begin{lemma}\label{lem:boundederror_sample}
Fix any $\delta \geq 0$, and define the event
\[\Ecal_\delta = \left\{\sum_i \left[\log\left(\frac{\Pj(\X_{ij}\mid \X_{i,-j})\Phj(\Xko_{ij}\mid \X_{i,-j})}{\Phj(\X_{ij}\mid \X_{i,-j})\Pj(\Xko_{ij}\mid \X_{i,-j})}\right)\right]^2\leq n\delta^2\text{ for all $j$}\right\}.\]
Then
\[\PP{\max_{j=1,\dots,p}\kl_j\leq \frac{n\delta^2}{2} +  2\delta\sqrt{n\log(p)}} \geq  1- \frac{1}{p} - \PP{(\Ecal_\delta)^c}.\]
\end{lemma}
\noindent In order to prove \lemref{boundederror}, then, we simply
observe that if the universal bound~\eqnref{delta_bound_wp1} holds for the likelihood ratios,
then the event $\Ecal_\delta$ occurs with probability 1.

Now we prove the general result, \lemref{boundederror_sample}.
Fix any $j$.
Suppose that we condition on $\X_{-j},\Xko_{-j}$,
and on the unordered pair $\{\X_{ij},\Xko_{ij}\} = \{a_{ij},b_{ij}\}$ for each $i$---that is, after observing the unlabeled pair, we arbitrarily label them as $a$ and $b$. Write  $a_j = (a_{1j},\dots,a_{nj})$ and same
for $b_j$. Let
 $C_{ij}=0$ if $a_{ij}=b_{ij}$, and otherwise
let 
\[C_{ij} := \begin{cases}+1,&\text{ if }(\X_{ij},\Xko_{ij}) =(a_{ij},b_{ij}),\\
-1,&\text{ if }(\X_{ij},\Xko_{ij}) =(b_{ij},a_{ij}).\end{cases}\]
Then we have
\begin{multline*}
\kl_j= \sum_i \log\left(\frac{\Pj (\X_{ij}\mid \X_{i,-j})\cdot \Phj(\Xko_{ij}\mid \X_{i,-j})}{\Phj(\X_{ij}\mid \X_{i,-j})\cdot \Pj (\Xko_{ij}\mid \X_{i,-j})}  \right)\\
= \sum_i 
C_{ij}\log\left(\frac{\Pj (a_{ij}\mid \X_{i,-j})\cdot \Phj(b_{ij}\mid \X_{i,-j})}{\Phj(a_{ij}\mid \X_{i,-j})\cdot \Pj (b_{ij}\mid \X_{i,-j})} \right)
\eqqcolon \sum_i C_{ij}\kl_{ij}.
\end{multline*}
By \lemref{approx_pairwise}, for each $i$ with $a_{ij}\neq b_{ij}$ we have
\begin{align}\label{eqn:Cij_prob}\frac{\PPst{C_{ij}=+1}{a_j,b_j,\X_{-j},\Xko_{-j}}}{\PPst{C_{ij}=-1}{a_j,b_j,\X_{-j},\Xko_{-j}}} \nonumber 
&=\frac{\PPst{(\X_{ij},\Xko_{ij}) = (a_{ij},b_{ij})}{a_j,b_j,\X_{-j},\Xko_{-j}}}{\PPst{(\X_{ij},\Xko_{ij}) = (b_{ij},a_{ij})}{a_j,b_j,\X_{-j},\Xko_{-j}}} \nonumber \\&= \frac{\Pj (a_{ij}\mid\X_{i,-j})\Phj(b_{ij}\mid \X_{i,-j})}{\Phj(a_{ij}\mid\X_{i,-j})\Pj (b_{ij}\mid \X_{i,-j})} = e^{\kl_{ij}}.
\end{align}
Note that this binary outcome is independent for each $i$.
From this point on we treat $\X_{-j},\Xko_{-j},a_{j},b_{j}$ as fixed (where $a_j = (a_{1j},\dots,a_{nj})$ and same
for $b_j$), and only the $C_{ij}$'s as random.
Since $\kl_{ij}$ depends only on $\X_{-j},\Xko_{-j},a_{j},b_{j}$  (i.e.~on the variables that we are conditioning on),
and is therefore treated as fixed, while $|C_{ij}|\leq 1$ by definition, we see that, writing $\mu_j = \EEst{\kl_j}{\X_{-j},\Xko_{-j},a_{j},b_{j}}$,
\[\PPst{\kl_j-\mu_j\geq 2\sqrt{\log(p)} \sqrt{\sum_i (\kl_{ij})^2}}{\X_{-j},\Xko_{-j},a_{j},b_{j}}\leq \frac{1}{p^2}\]
by Hoeffding's inequality.
Next we work with the conditional expectation of $\kl_j$.
For any $i$ with $a_{ij}\neq b_{ij}$, we use~\eqnref{Cij_prob} to calculate
\[
\left|\EEst{C_{ij}}{\X_{-j},\Xko_{-j},a_{j},b_{j}}\right|
= \left|\frac{e^{\kl_{ij}} - 1}{e^{\kl_{ij}}+1}\right|\leq \frac{{|\kl_{ij}|}}{2}.\]
Then
\[\left|\EEst{\kl_j}{\X_{-j},\Xko_{-j},a_{j},b_{j}}\right| = \left|\sum_i \EEst{C_{ij}}{\X_{-j},\Xko_{-j},a_{j},b_{j}}\cdot \kl_{ij}\right| \\
\leq \frac{1}{2}\sum_i (\kl_{ij})^2.\]
Therefore, combining everything,
\[\PPst{\kl_j\geq \frac{1}{2}\sum_i (\kl_{ij})^2 + 2\sqrt{\log(p)} \sqrt{\sum_i (\kl_{ij})^2}}{\X_{-j},\Xko_{-j},a_{j},b_{j}}\leq \frac{1}{p^2}.\]
Now, under the event $\Ecal_\delta$ we must have $\sum_i (\kl_{ij})^2\leq n\delta^2$,
and so we can write
\[\PPst{\kl_j\cdot\One{\Ecal_\delta}\geq \frac{n\delta^2}{2} + 2\delta\sqrt{n\log(p)}}{\X_{-j},\Xko_{-j},a_{j},b_{j}}\leq \frac{1}{p^2} .\]
Marginalizing over all the conditioned variables, and taking a union bound over all $j$, we have proved that
\[\PP{\max_{j=1,\dots,p}\kl_j\cdot\One{\Ecal_\delta}\geq \frac{n\delta^2}{2} + 2\delta\sqrt{n\log(p)}}\leq \frac{1}{p} .\]
This proves the lemma.

\subsection{Proof of \lemref{gaussian_pairwise}}

Fix any feature index $j$, and consider any distribution $D^{(j)}$ on $\R^p$ with $j$th conditional equal to $\Phj$, as defined in~\eqnref{gaussian_Qj}.
For simplicity, from this point on, we will perform calculations treating $D^{(j)}$ as a joint density, but the result is valid without this assumption. 
Drawing $X\sim D^{(j)}$ and $\bigxko\mid X \sim \Pt(\cdot|X)$, then the joint density of $(X,\bigxko)$ is given by
\[D^{(j)}(x)\cdot \Pt(\xko\mid x) \\
=
\underbrace{D^{(j)}_{-j}(x_{-j})}_{\textnormal{Term 1}}\cdot
\underbrace{\left(\frac{ \Phj(x_j\mid x_{-j}) }{
      \exp\left\{-\frac{1}{2}x^\top \Thetah
        x\right\}}\right)}_{\textnormal{Term 2}} \cdot
\underbrace{\left( \Pt(\xko\mid x)\cdot
    \exp\left\{-\frac{1}{2}x^\top\Thetah
      x\right\}\right)}_{\textnormal{Term 3}},\]
where $D^{(j)}_{-j}$ is the marginal distribution of $X_{-j}$ under
the joint distribution $X\sim D^{(j)}$. In order to check
that $(X_j,\bigxko_j,X_{-j},\bigxko_{-j})\eqd
(\bigxko_j,X_j,X_{-j},\bigxko_{-j})$
under this distribution, we therefore need to check that this joint
density is exchangeable in the variables $x_j$ and $\xko_j$; that is,
swapping $x_j$ and $\xko_j$ does not change the value of the joint
density $D^{(j)}(x)\cdot \Pt(\xko\mid x)$.  We check this by
considering each of the three terms separately. Term 1 clearly does
not depend on either $x_j$ or $\xko_j$. Next, using the calculation of
$\Phj$ in~\eqnref{gaussian_Qj}, we can simplify Term 2 to obtain
\begin{multline*}
\textnormal{Term 2} \propto \exp\left\{ - \frac{1}{2/\Thetah_{jj}}\left(x_j +x_{-j}^\top \Thetah_{-j,j}/\Thetah_{jj}\right)^2 +  \frac{1}{2}x^\top \Thetah x\right\}\\
= \exp\left\{  \frac{1}{2}x_{-j}^\top \left(\Thetah_{-j,-j} - \frac{\Thetah_{-j,j}\Thetah_{-j,j}^\top}{\Thetah_{jj}} \right)x_{-j}\right\},
\end{multline*}
which also does not depend on either $x_j$ or $\xko_j$. Finally, Term 3 is exchangeable in the pair $x_j,\xko_j$ by the construction of the knockoff distribution $\Pt$.
More concretely, using the definition of $\Pt$ given in~\eqnref{draw_knockoffs_Gaussian}, we can calculate
\begin{align*}
\textnormal{Term 3} 
&\propto\exp\left\{-\frac{1}{2} \big(\xko - (\ident - D \Thetah)x\big)^\top (2D-D\Thetah D)^{-1}\big(\xko - (\ident - D\Thetah)x\big) -\frac{1}{2}x^\top\Thetah x\right\} \\
&=\exp\left\{ - \frac{1}{2} (x+\xko)^\top  (2D-D\Thetah D)^{-1} (x+\xko) + x^\top D^{-1} \xko\right\},
\end{align*}
which is clearly exchangeable in the pair $x_j,\xko_j$ (note that the exchangeability of $x_j,\xko_j$ in the term $x^\top D^{-1} \xko$
follows from the fact that $D$ is a diagonal matrix).

\subsection{Proof of \lemref{Gaussian}}
We will apply \lemref{boundederror_sample} to prove this result.
We first recall the  conditional distributions $\Pj$ for the joint distribution $\PX = \normal_p(0,\Theta)^{-1}$,
which can be computed as
\[\Pj(\cdot|x_{-j}) = \normal\left(x_{-j}^\top \left(-\Theta_{-j,j}/\Theta_{jj}\right), 1/\Theta_{jj}\right),\]
 and the conditionals  $\Phj$, calculated earlier in~\eqnref{gaussian_Qj} as \[
\Phj(\cdot|x_{-j}) = \normal\left(x_{-j}^\top \left(-\Thetah_{-j,j}/\Thetah_{jj}\right), 1/\Thetah_{jj}\right).\]
Then we can calculate
\begin{multline*}\sum_i \left[\log\left(\frac{\Pj(\X_{ij}\mid \X_{i,-j})\Phj(\Xko_{ij}\mid \X_{i,-j})}{\Phj(\X_{ij}\mid \X_{i,-j})\Pj(\Xko_{ij}\mid \X_{i,-j})}\right) \right]^2\\
=\sum_i \bigg[-(\X_{ij}-\Xko_{ij})\cdot \frac{\Thetah_{jj}-\Theta_{jj}}{2} +\X_{i*}^\top\big(\Thetah_j-\Theta_j\big)\bigg]^2\cdot \Big[\X_{ij}-\Xko_{ij}\Big]^2\\
\leq \frac{1}{2}\sum_i \bigg[\underbrace{-(\X_{ij}-\Xko_{ij})\cdot \frac{\Thetah_{jj}-\Theta_{jj}}{2} +\X_{i*}^\top\big(\Thetah_j-\Theta_j\big)}_{\sim \normal(0,v_j^2) \text{ for each $i$}}\bigg]^4 + \frac{1}{2}\sum_i  \Big[\underbrace{\X_{ij}-\Xko_{ij}}_{\sim \normal(0,w_j^2)\text{ for each $i$}}\Big]^4.\end{multline*}
Using standard tail bounds on Gaussian and $\chi^2$ random variables, and computing the variances $v_j^2$ and $w_j^2$,
after some calculations we can show that the quantity above is bounded as
\begin{align*}
\sum_i &\left[\log\left(\frac{\Pj(\X_{ij}\mid \X_{i,-j})\Phj(\Xko_{ij}\mid \X_{i,-j})}{\Phj(\X_{ij}\mid \X_{i,-j})\Pj(\Xko_{ij}\mid \X_{i,-j})}\right) \right]^2 \\
&\hspace{2cm}\leq 4\left[\left(\frac{\delta_\Theta}{1-\delta_\Theta}\right)^2 + \left(\frac{\delta_\Theta}{1-\delta_\Theta}\right)^4\right]\cdot \left(\sqrt{n} + 2\sqrt{\log(np)}\right)^2,\end{align*}
with probability at least $1-\frac{1}{p}$, and therefore, $\PP{\Ecal_\delta}\geq 1 - \frac{1}{p}$ when we take
\[\delta = 2\sqrt{\left(\frac{\delta_\Theta}{1-\delta_\Theta}\right)^2 + \left(\frac{\delta_\Theta}{1-\delta_\Theta}\right)^4}\cdot \left(1 + 2\sqrt{\frac{\log(np)}{n}}\right) =2\delta_\Theta \cdot (1+o(1)) ,\]
where the last step holds as long as $\frac{\log(p)}{n}=o(1)$ and $\delta_\Theta = o(1)$.
Applying \lemref{boundederror_sample} then proves that
\[\PP{\max_{j=1,\dots,p}\kl_j\leq \frac{n\delta^2}{2} +  2\delta\sqrt{n\log(p)}} \geq  1- \frac{2}{p}.\]
Assuming this upper bound on the $\kl_j$'s is bounded by a constant, the dominant term is therefore $2\delta\sqrt{n\log(p)}$,
which proves the lemma.

\subsection{Proof of \lemref{Tj}}
First, recall that $T=T(W)$ is defined as follows:
\[T = \min\bigg\{ t \geq \eps(W) : \underbrace{\frac{ c + \sum_{\ell=1}^p \One{W_\ell \leq -t}}{\sum_{\ell=1}^p \One{W_\ell\geq t}}}_{\eqqcolon f(W,t)} \leq q\bigg\},\]
where $\eps (W)>0$ is chosen to be the smallest magnitude of the $W$ statistics, i.e.~$\eps(W) = \min\{|W_\ell| : |W_\ell|>0\}$, and where $c=0$ for knockoff or $c=1$ for knockoff+.
Next, define
\[W^j := (W_1,\dots,W_{j-1},|W_j|,W_{j+1},\dots,W_p)\]
and similarly
\[W^k := (W_1,\dots,W_{k-1},|W_k|,W_{k+1},\dots,W_p),\]
so that $T_j = T(W^j)$ and $T_k = T(W^k)$. Note that $|W^j| = |W^k| = |W|$, and so $\eps(W^j) = \eps(W^k) = \eps(W)$ since $\eps(W)$ depends on $W$ only through $|W|$.

Without loss of generality, assume $T_j\leq T_k$, so that by assumption we have $W_j\leq -T_j$ and $W_k\leq -T_j$.
Consider
\begin{equation*}
f(W^k,T_j) = \frac{ c + \sum_{\ell=1}^p \One{W^k_\ell \leq -T_j}}{\sum_{\ell=1}^p \One{W^k_\ell\geq T_j}}.
\end{equation*}
We will next rewrite the numerator and denominator. Beginning with the numerator, we have
\begin{multline*}\sum_{\ell=1}^p \One{W^k_\ell \leq -T_j} = \sum_{\ell=1}^p \One{W^j_\ell \leq -T_j}+{}\\
 \One{W^k_j \leq -T_j} - \One{W^j_j\leq -T_j} + \One{W^k_k\leq -T_j} - \One{W^j_k \leq -T_j}  \\
 = \sum_{\ell=1}^p \One{W^j_\ell \leq -T_j}+ (1-0 + 0 - 1)=  \sum_{\ell=1}^p \One{W^j_\ell \leq -T_j},\end{multline*}
where the first step holds since $W^j$ and $W^k$ differ only on entries $j,k$, while the second step holds because 
we know from our assumptions and definitions that $W^k_j = W_j \leq -T_j$,  $W^j_j = |W_j| \geq T_j$,
$W^k_k = |W_k| \geq T_j$, and $W^j_k = W_k\leq -T_j$.
Similarly, for the denominator, we have
\begin{multline*}\sum_{\ell=1}^p \One{W^k_\ell \geq T_j} = \sum_{\ell=1}^p \One{W^j_\ell \geq T_j}+{}\\
 \One{W^k_j \geq T_j} - \One{W^j_j\geq T_j} + \One{W^k_k\geq T_j} - \One{W^j_k \geq T_j}  \\
 = \sum_{\ell=1}^p \One{W^j_\ell \geq T_j}+ (0 - 1 + 1 - 0)=  \sum_{\ell=1}^p \One{W^j_\ell \geq T_j}.\end{multline*}
Therefore,
\[f(W^k,T_j) = \frac{ c + \sum_{\ell=1}^p \One{W^k_\ell \leq -T_j}}{\sum_{\ell=1}^p \One{W^k_\ell\geq T_j}}
 =  \frac{ c + \sum_{\ell=1}^p \One{W^j_\ell \leq -T_j}}{\sum_{\ell=1}^p \One{W^j_\ell\geq T_j}} \\
 = f(W^j,T_j) \leq q,\]
where the last step holds by definition of $T_j$. Therefore, since $T_j\geq \eps(W^j) = \eps(W^k)$,
we see from the definition of $T_k$ that we must have $T_k\leq T_j$. This proves that $T_j=T_k$, as desired.

\end{document}